\documentclass{elsart}

\usepackage[square,comma]{natbib}
\usepackage{graphicx}
\usepackage{pxfonts}
\usepackage{lineno}

\usepackage{amssymb}

\journal{}

\begin{document}

\thispagestyle{empty}
\begin{Large}
\textbf{DEUTSCHES ELEKTRONEN-SYNCHROTRON}

\textbf{\large{Ein Forschungszentrum der Helmholtz-Gemeinschaft}\\}
\end{Large}

DESY 12-121

July 2012

\begin{eqnarray}
\nonumber &&\cr \nonumber && \cr \nonumber &&\cr
\end{eqnarray}
\begin{eqnarray}
\nonumber
\end{eqnarray}
\begin{center}
\begin{Large}
\textbf{Extension of self-seeding scheme with single crystal
monochromator to lower energy $<$ 5 keV as a way to generate
multi-TW scale pulses  at the European XFEL}
\end{Large}
\begin{eqnarray}
\nonumber &&\cr \nonumber && \cr
\end{eqnarray}

\begin{large}
Gianluca Geloni,
\end{large}
\textsl{\\European XFEL GmbH, Hamburg}
\begin{large}

Vitali Kocharyan and Evgeni Saldin
\end{large}
\textsl{\\Deutsches Elektronen-Synchrotron DESY, Hamburg}
\begin{eqnarray}
\nonumber
\end{eqnarray}
\begin{eqnarray}
\nonumber
\end{eqnarray}
ISSN 0418-9833
\begin{eqnarray}
\nonumber
\end{eqnarray}
\begin{large}
\textbf{NOTKESTRASSE 85 - 22607 HAMBURG}
\end{large}
\end{center}
\clearpage
\newpage

\begin{frontmatter}



\title{Extension of self-seeding scheme with single crystal
monochromator to lower energy $<$ 5 keV as a way to generate
multi-TW scale pulses  at the European XFEL}


\author[XFEL]{Gianluca Geloni\thanksref{corr},}
\thanks[corr]{Corresponding Author. E-mail address: gianluca.geloni@xfel.eu}
\author[DESY]{Vitali Kocharyan}
\author[DESY]{and Evgeni Saldin}

\address[XFEL]{European XFEL GmbH, Hamburg, Germany}
\address[DESY]{Deutsches Elektronen-Synchrotron (DESY), Hamburg,
Germany}

\begin{abstract}
We propose a use of the self-seeding scheme with single crystal
monochromator to produce high power, fully-coherent pulses for
applications at a dedicated bio-imaging beamline at the European
X-ray FEL in the photon energy range between 3.5 keV and 5 keV. We
exploit the C(111) Bragg reflection ($\pi$-polarization) in diamond
crystals with a thickness of 0.1 mm, and we show that, by tapering
the 40 cells of the SASE3 type undulator the FEL power can reach up
to 2 TW in the entire photon energy range. The present design
assumes the use of a nominal electron bunch with charge 0.1 nC  at
nominal electron beam energy 17.5 GeV. The main application of the
scheme proposed in this work is for single shot imaging of
individual protein molecules.
\end{abstract}

%
%
%
\end{frontmatter}




\section{\label{sec:intro} Introduction}

Despite the unprecedented increase in peak power of X-ray pulses
from SASE X-ray FELs, some applications including imaging of complex
molecules like proteins and other biologically interesting
structures may still require higher photon flux (see, among others,
\cite{HAJD}-\cite{CHAP}). The most promising way to extract more FEL
power than that at saturation is by tapering the magnetic field of
the undulator \cite{TAP1}-\cite{TAP4}. Also, a significant increase
in power is achievable by starting the FEL process from
monochromatic seed rather than from noise \cite{FAWL}-\cite{WANG}.
Self-seeding is a promising approach to significantly narrow the
SASE bandwidth and to produce nearly transform-limited pulses
\cite{SELF}-\cite{OURY5b}. The combination of self-seeding and
tapering techniques would allow to meet the desired TW-scale output
power for bio-imaging applications \cite{OURY3}-\cite{LAST}.

We recently proposed a study for a possible dedicated bio-imaging
beamline at the European XFEL \cite{OBIO}. In that concept we
suggested that the use of a single crystal self-seeding scheme would
allow to deliver nearly transform-limited TW-scale X-ray pulses in
the photon energy range between 8 keV and 13 keV. However, potential
users of such a bio-imaging beamline mainly wish to investigate
their samples in the energy range between 3 keV and 5 keV, where the
diffraction signal is stronger. Finding a solution suitable for this
spectral range is major challenge for self-seeding designers. In
fact, due to high absorption, both single crystal monochromators and
grating monochromators have a low throughput in the energy range
between 3 keV and 5 keV. In \cite{OBIO} we proposed a method to get
around this obstacle, which is based in essence on a fresh bunch
technique \cite{BZVI} and exploits a conservative design of a
self-seeding setup based on grating monochromator \cite{FENG,FENG2}
in the photon energy range between 0.3 keV and 1.7 keV.

In this work we propose an alternative possibility to provide
bio-imaging capabilities in the photon energy range between 3 keV
and 5 keV, based on an extension of the original design of the
self-seeding scheme with single crystal monochromator down to 3 keV.
We still suggest to use a diamond crystal with thickness 0.1 mm.
However, we consider a symmetric C(111) Bragg reflection
($\pi$-polarization). This reflection will allow to cover the photon
energy range from 3.5 keV to 5 keV. We demonstrate that the
previously mentioned drawback of a low throughput can be overcome by
enabling a cascade self-seeding scheme \cite{OURY2}.

In its simplest configuration, a self-seeded XFEL consists of an
input undulator and an output undulator separated by a  single
crystal monochromator. At photon energies below 5 keV absorption in
crystals is very high and the simplest configuration, based on two
undulators, is not optimal. A possible extension is to use a setup
with three or more undulators separated by monochromators. Each
cascade consists of an undulator acting as an amplifier, and of a
single crystal monochromator. The amplification-monochromatization
cascade scheme is distinguished, in performance, by a high
signal-to-noise ratio and by small electron beam perturbations at
the entrance of the output undulator. In this paper we study such
scheme, which consists of two parts: first, a a succession of two
amplification-monochromatization cascades and, second, an output
undulator in which the monochromatic seed is amplified up to 2 TW
power.

\section{Principle of cascade self-seeding technique based on the use of single
crystal monochromators}

\begin{figure}[tb]
\includegraphics[width=1.0\textwidth]{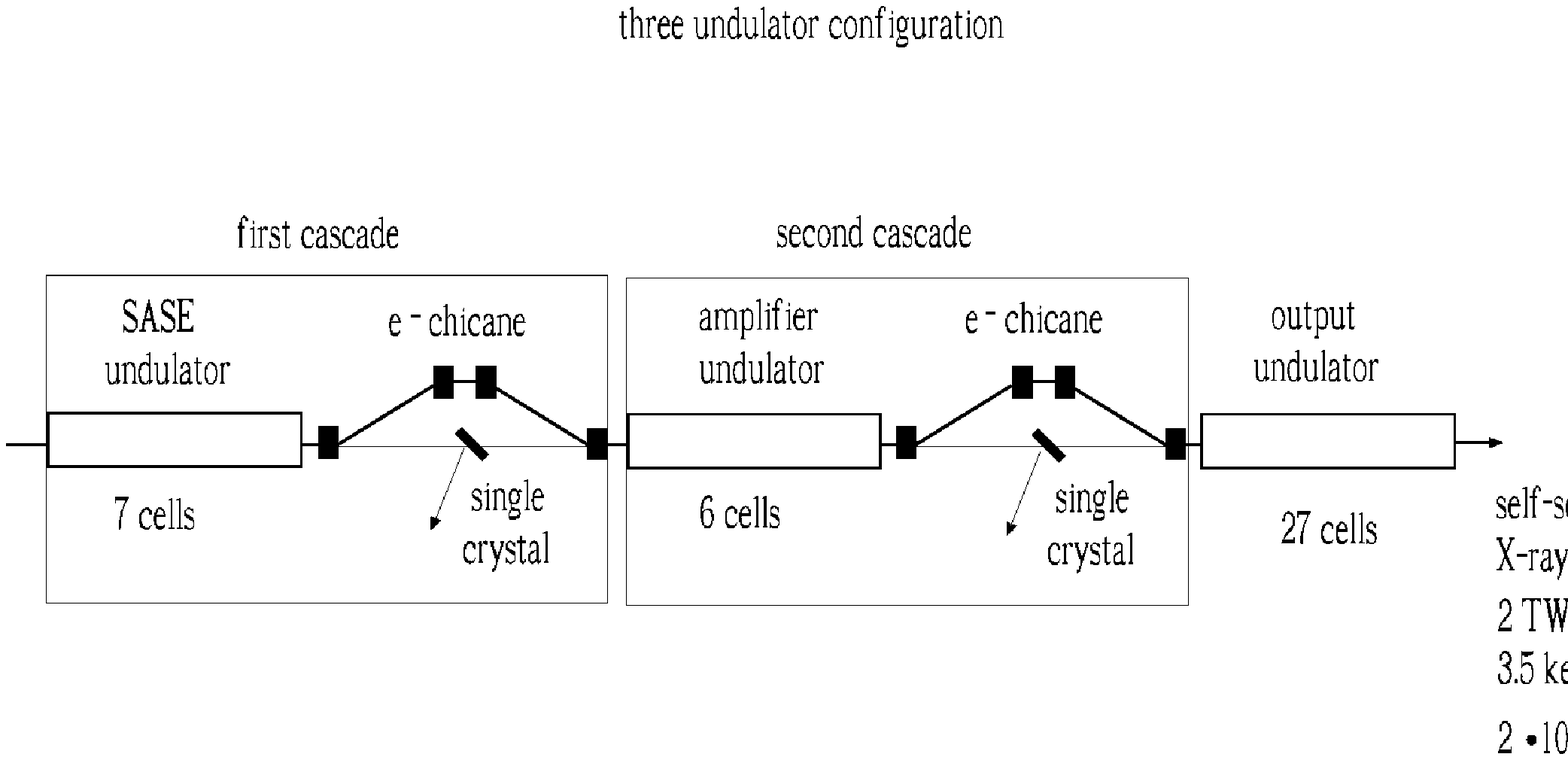}
\caption{Schematic of a two-cascade self-seeding scheme with single
crystal monochromators. This scheme holds a great promise as a
source of X-ray radiation in the 3.5 keV - 5 keV photon energy range
for applications such as single biomolecule imaging. }
\label{bio2f1}
\end{figure}

\begin{figure}[tb]
\includegraphics[width=0.5\textwidth]{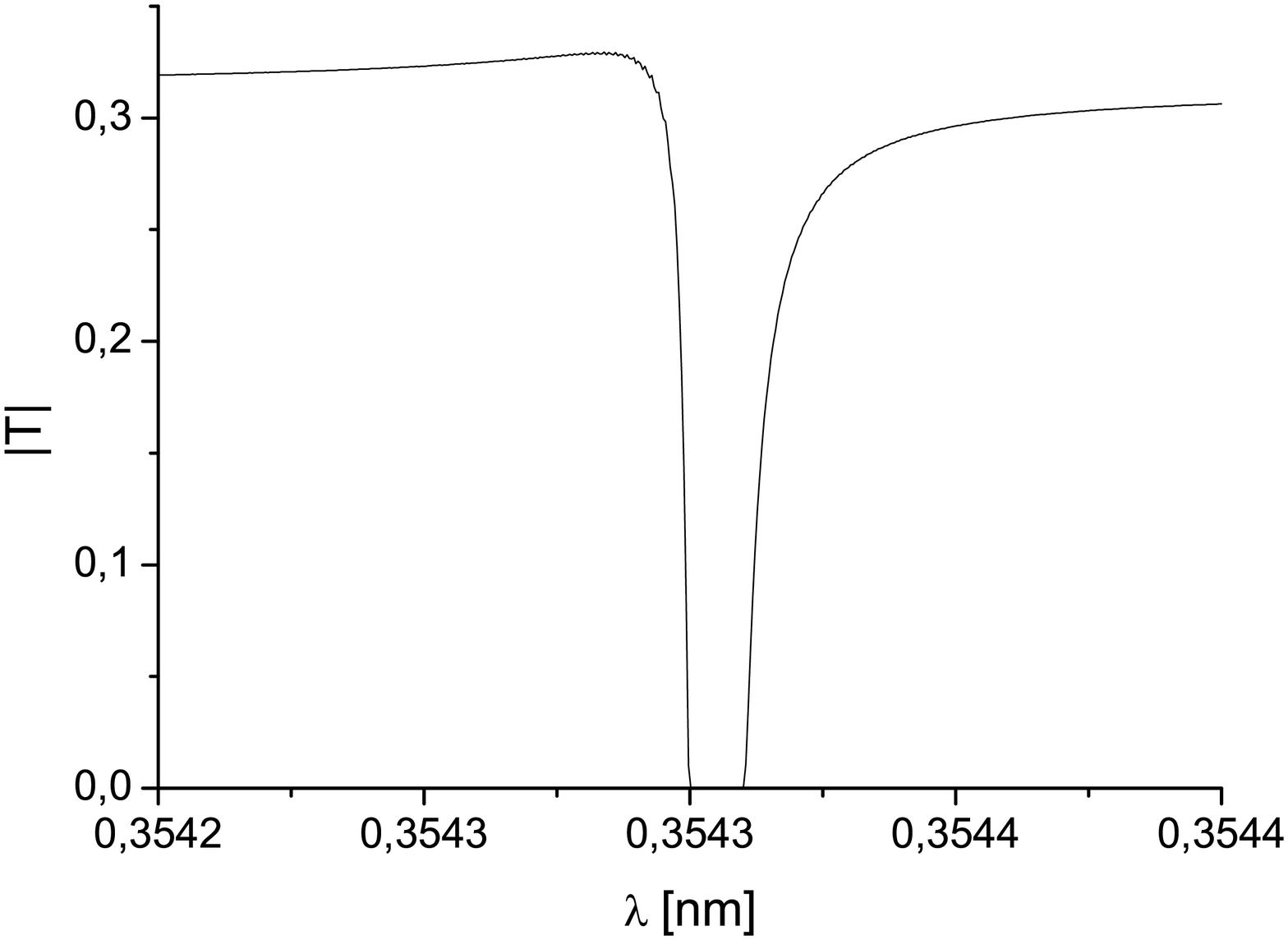}
\includegraphics[width=0.5\textwidth]{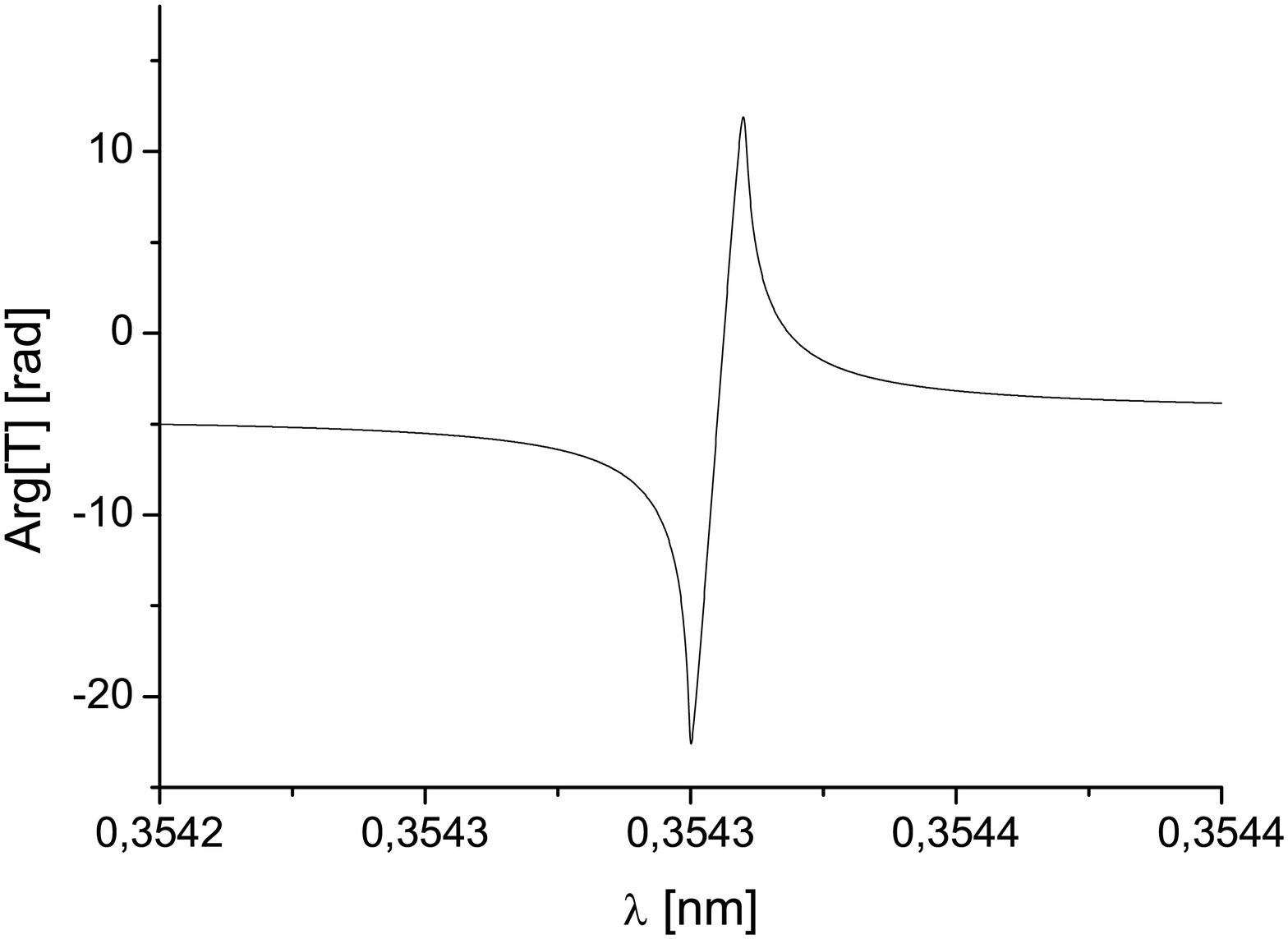}
\caption{Absolute value and phase of the transmission function
pertaining the C(111) forward diffraction of a $100~\mu$m-thick
diamond crystal ($\pi$-polarization). } \label{filter}
\end{figure}

\begin{figure}[tb]
\begin{center}
\includegraphics[width=0.7\textwidth]{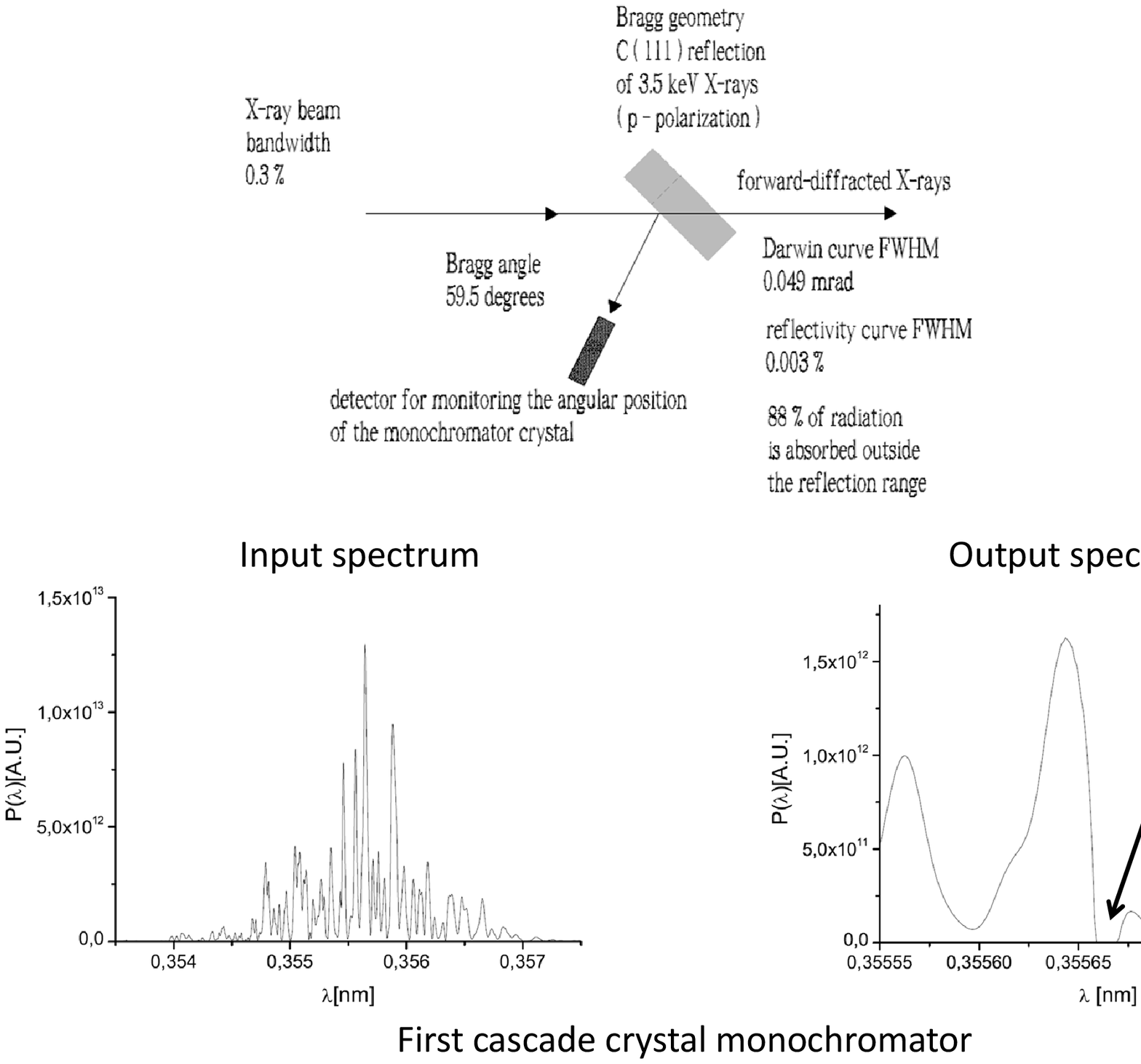}
\end{center}
\caption{Effect of the filtering process provided by the first
diamond crystal on the incident spectrum of the radiation. }
\label{bio2f3}
\end{figure}

\begin{figure}[tb]
\begin{center}
\includegraphics[width=0.75\textwidth]{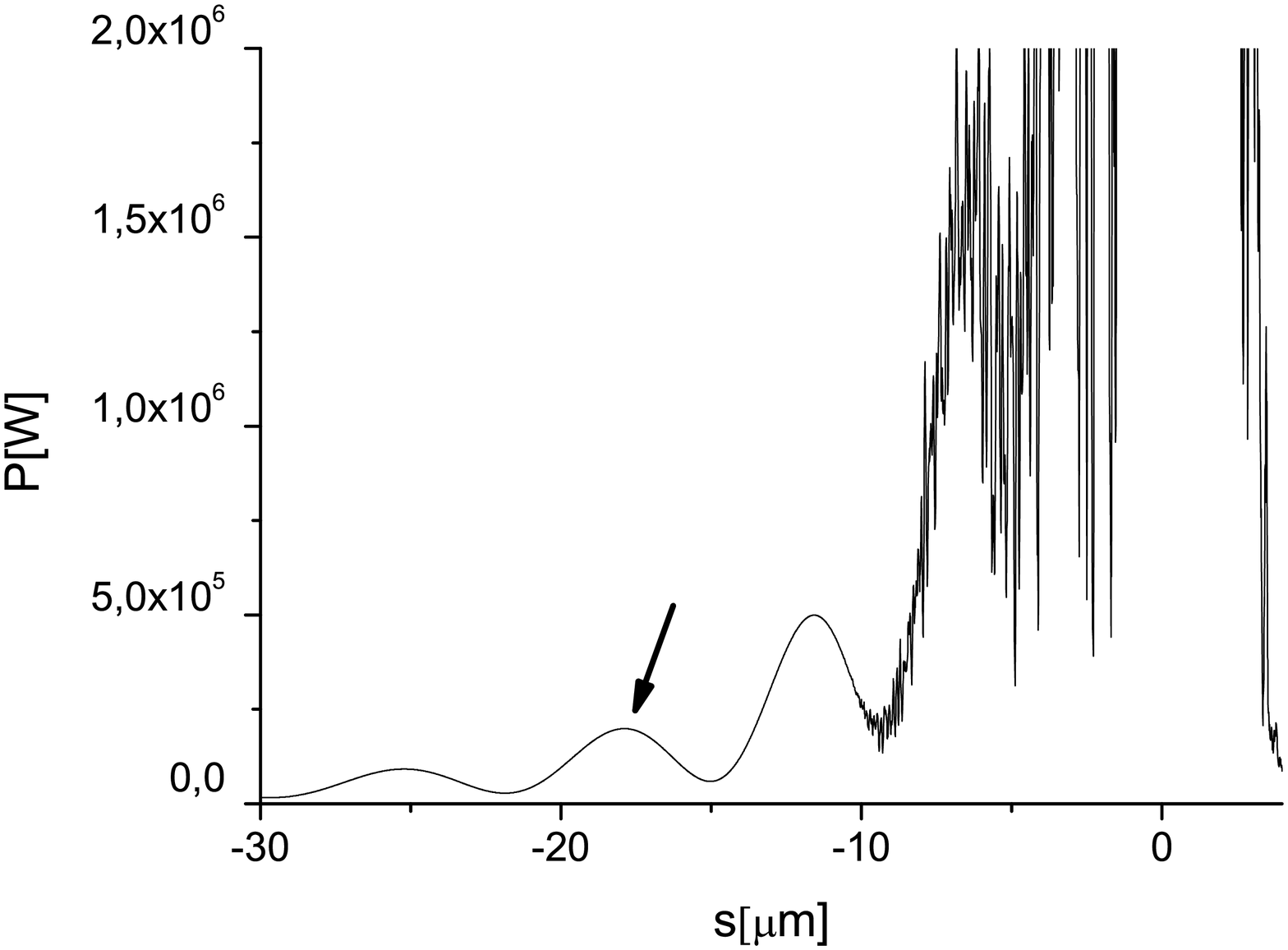}
\end{center}
\caption{Temporal shape of the seed signal from the first
self-seeding setup. The black arrow indicates the seeding region
used in this article.} \label{seed1}
\end{figure}

\begin{figure}[tb]
\includegraphics[width=0.5\textwidth]{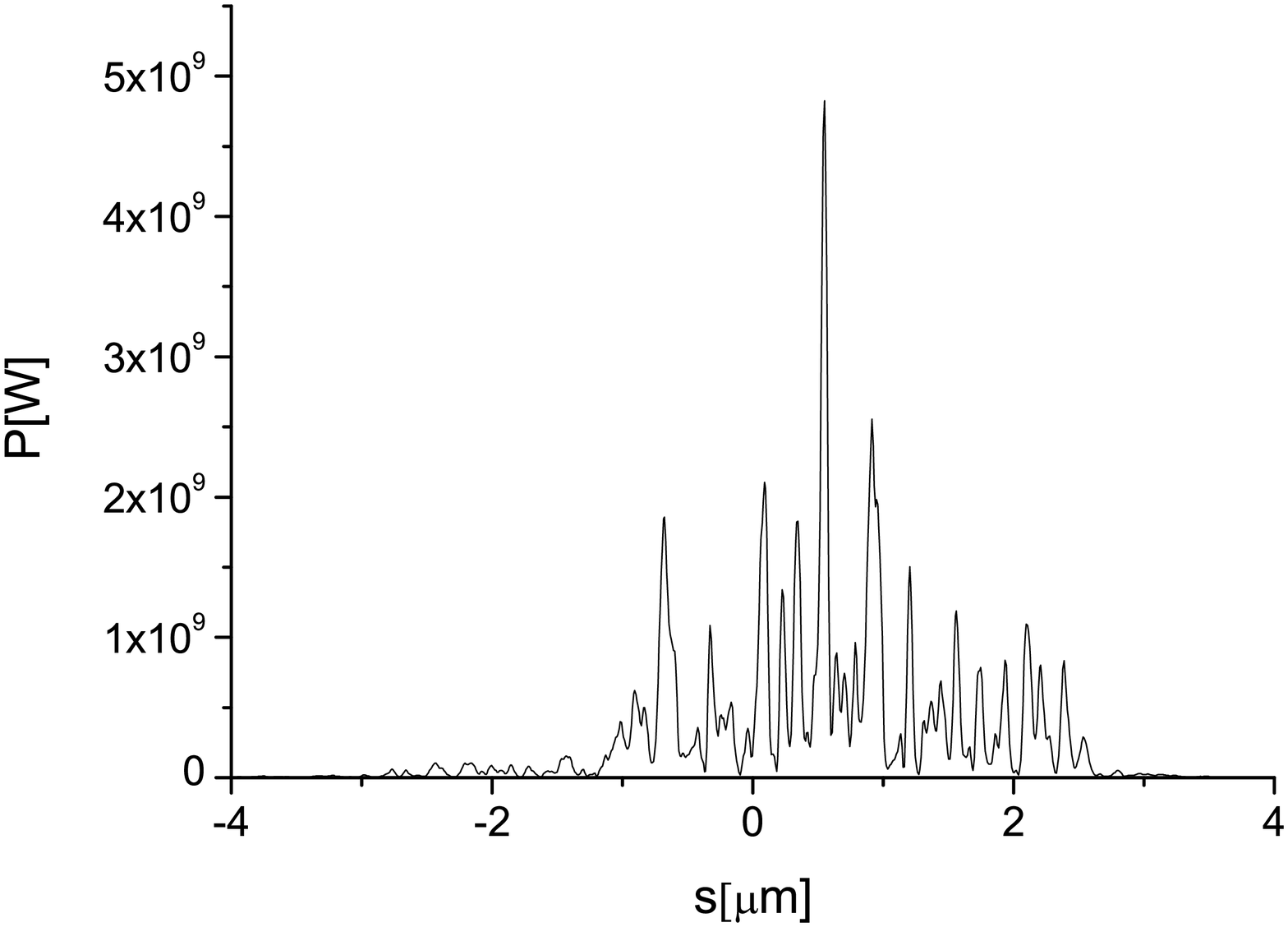}
\includegraphics[width=0.5\textwidth]{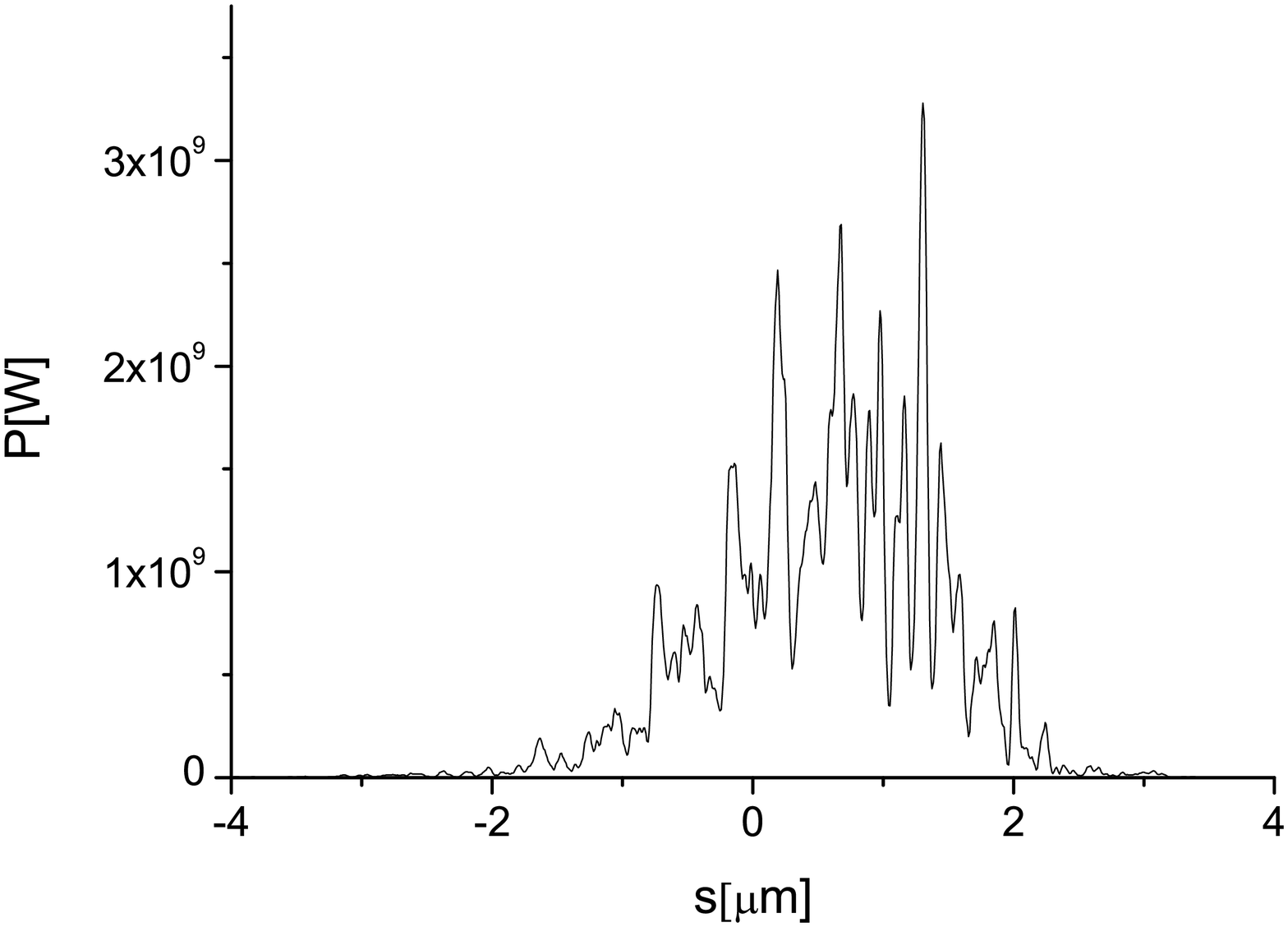}
\caption{Power before the first (left) and the second (right)
monochromator.} \label{powcomp}
\end{figure}

\begin{figure}[tb]
\begin{center}
\includegraphics[width=0.5\textwidth]{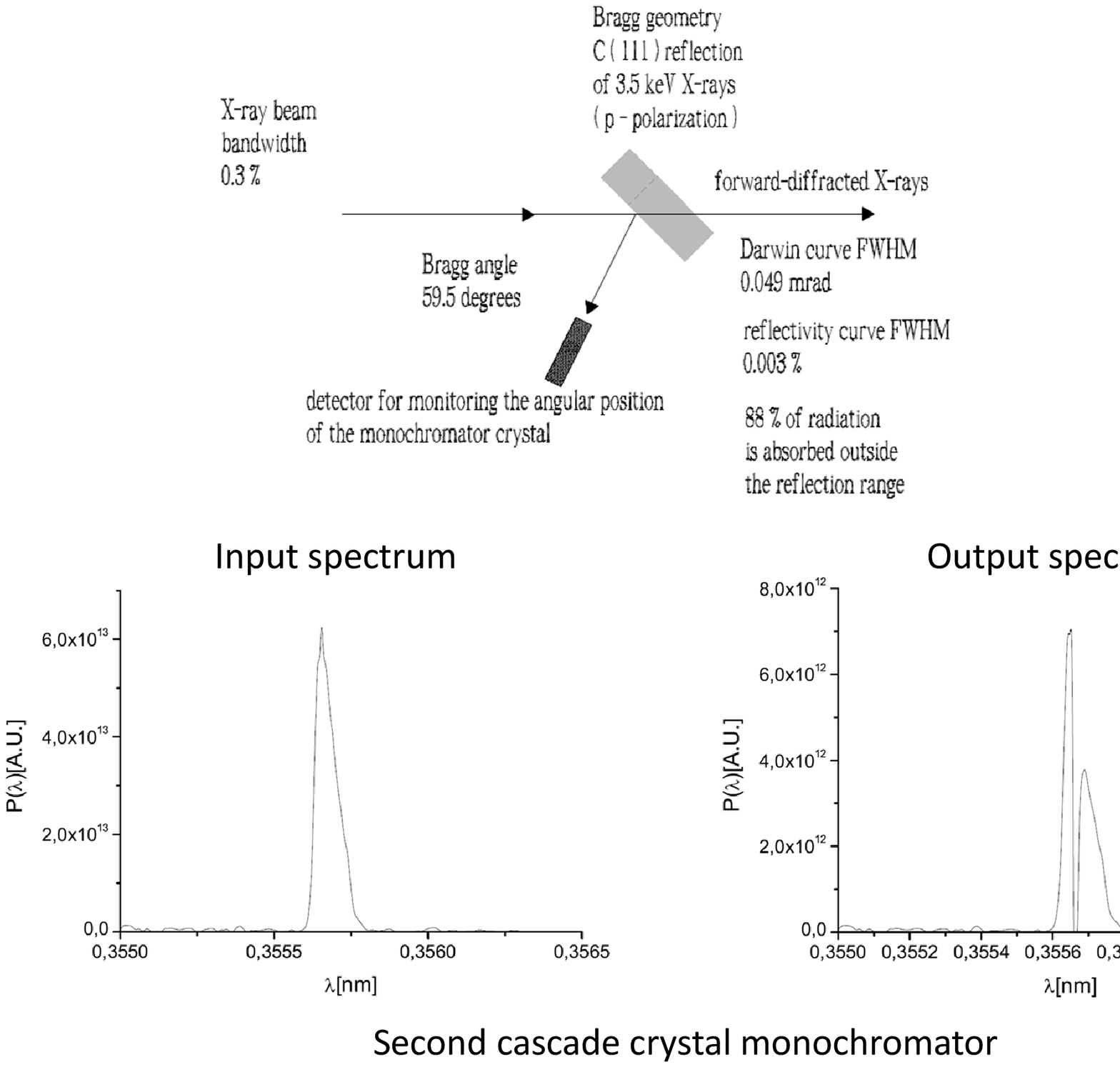}
\end{center}
\caption{Effect of the filtering process provided by the second
diamond crystal on the incident spectrum of the radiation.}
\label{bio2f4}
\end{figure}

\begin{figure}[tb]
\begin{center}
\includegraphics[width=0.75\textwidth]{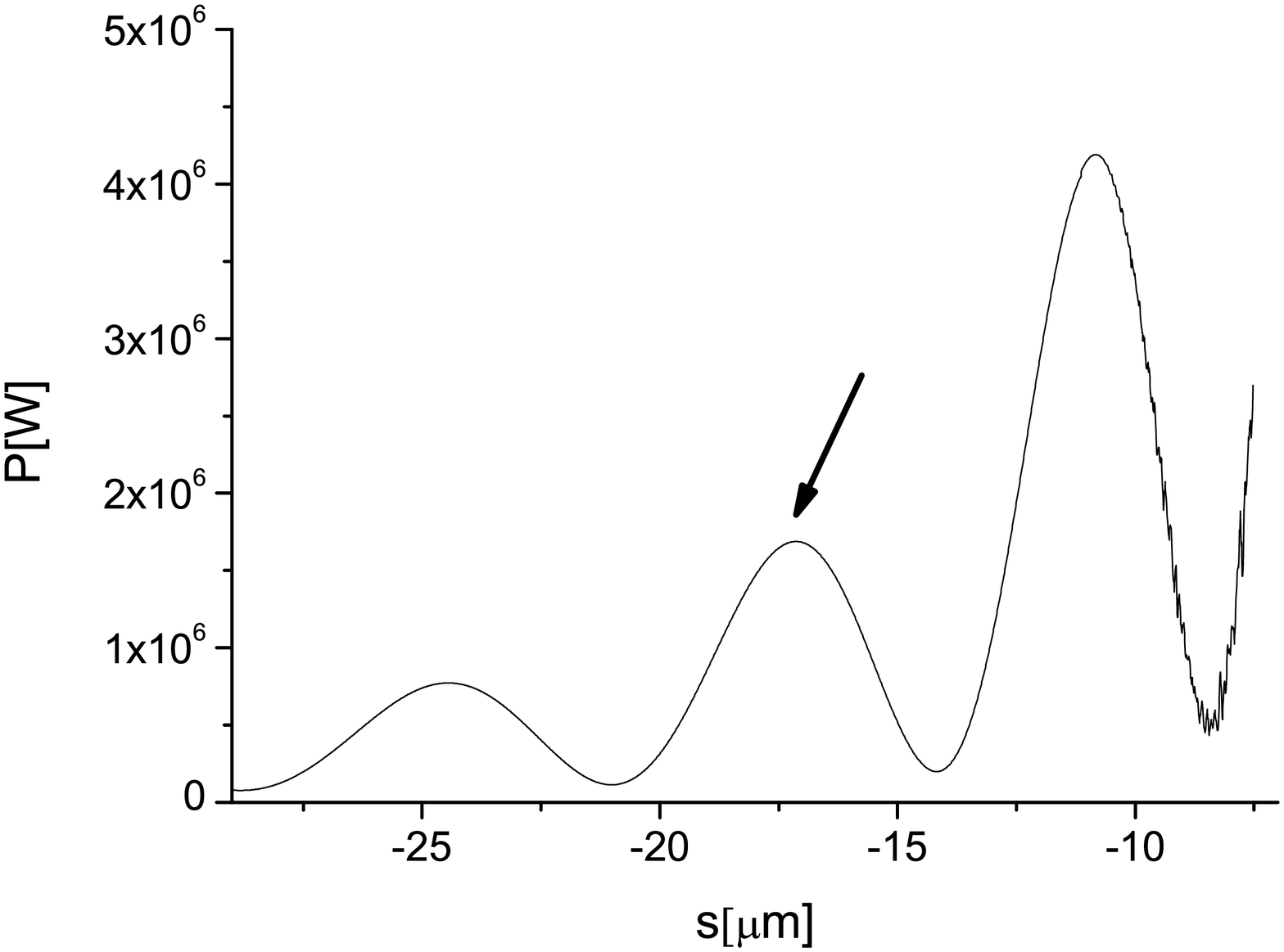}
\end{center}
\caption{Temporal shape of the seed signal from the second
self-seeding setup. The black arrow indicates the seeding region
used in this article.} \label{seed2}
\end{figure}

\begin{figure}[tb]
\includegraphics[width=0.5\textwidth]{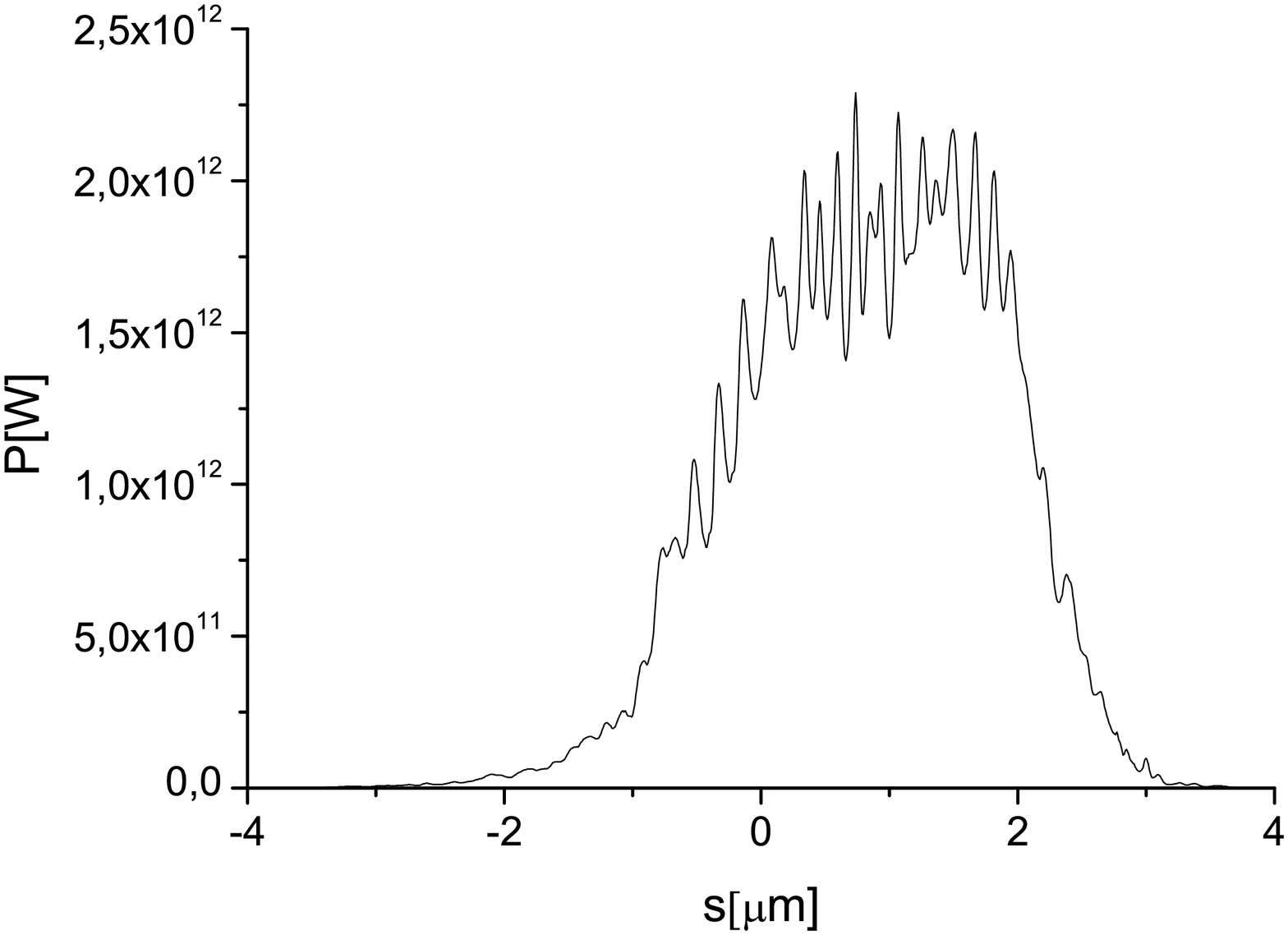}
\includegraphics[width=0.5\textwidth]{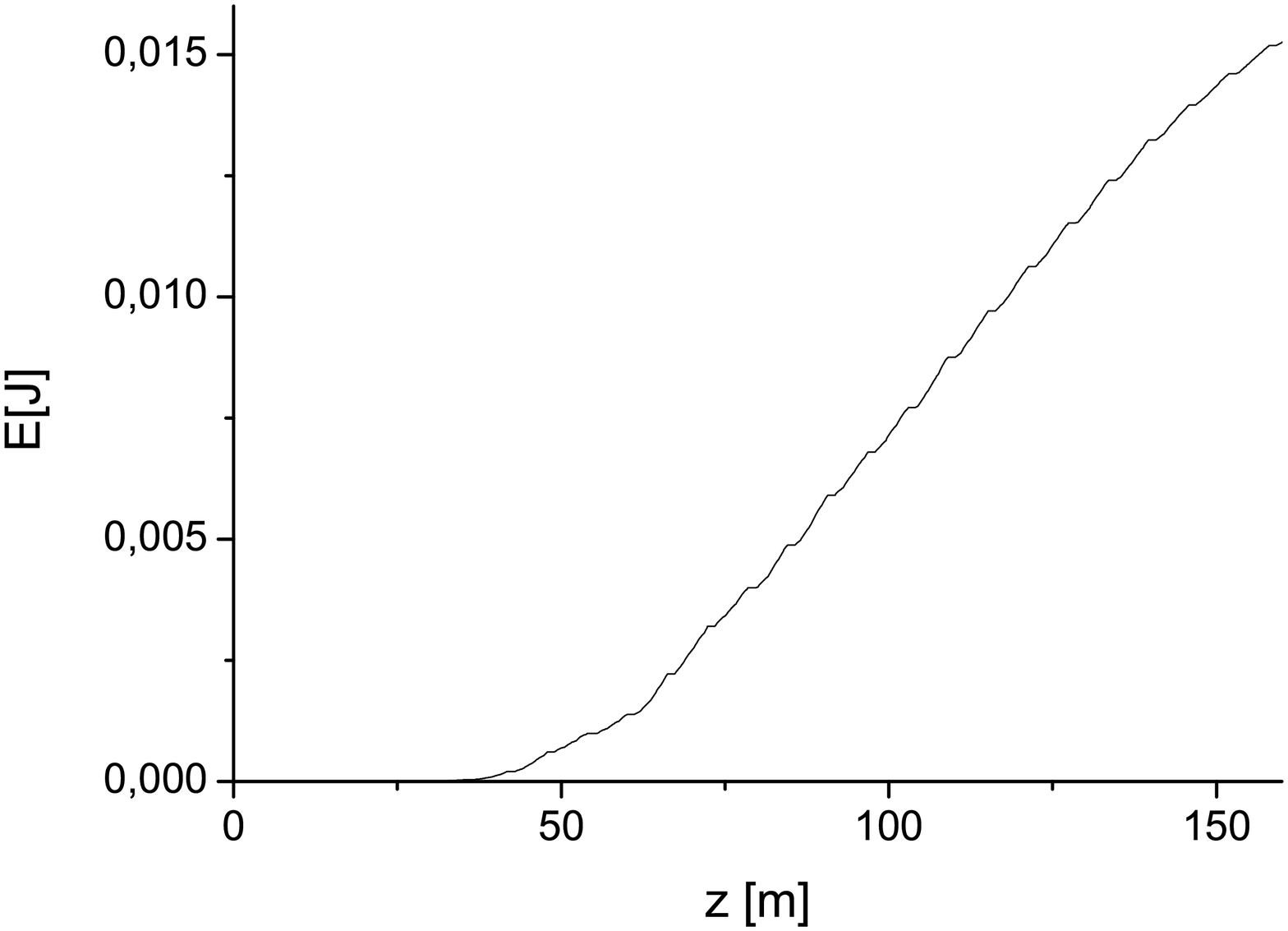}
\caption{(Left) The seeded FEL signal in the time domain after the
tapered output undulator. (Right) Energy of the seeded FEL pulse as
a function of the distance inside the output undulator.}
\label{outf}
\end{figure}

The self-seeding technique considered in this work is based on the
substitution of a single undulator module with a weak chicane and a
single crystal. Two cascades can be arranged sequentially as shown
in Fig. \ref{bio2f1}.

The first undulator in Fig. \ref{bio2f1} operates in the linear
high-gain regime starting from the shot-noise in the electron beam.
After the first undulator, the output SASE radiation passes through
the monochromator, which reduces the bandwidth to the desired value.
According to the wake monochromator principle, the SASE pulse coming
from the first undulator impinges on a crystal set for Bragg
diffraction.

The single crystal in Bragg geometry actually operates as a bandstop
filter for the transmitted X-ray SASE radiation pulse. The filter
transmission function in modulus and phase is shown in Fig.
\ref{filter}, where the central bandwidth can be tuned by properly
tilting the crystal. In Fig. \ref{bio2f3} we show the effect of the
filtering through the crystal on the incident spectrum of the
radiation. When the incident angle and the spectral contents of the
incoming beam satisfy the Bragg diffraction condition, the temporal
waveform of the transmitted radiation pulse shows a long
monochromatic wake, whose particular shape is linked to the shape of
the filter in the frequency domain. The overall duration of this
wake is inversely proportional to the bandwidth of the absorption
line in the transmittance spectrum, while the particular shape of
the wake, which in our case exhibits several oscillations on a
shorter temporal scale, is due to the particular characteristics of
the filter in the frequency domain. These characteristics are
calculated in the frequency domain with the help of the dynamical
theory of X-ray diffraction. In other words, all the physics
involved can be limited to the frequency domain. After this task is
accomplished, the particular shape of the wake can be simply
interpreted as a consequence of a Fourier transform. It should be
noted that, for the energy range under examination, between $3$ keV
and $5$ keV, it is useful to consider the $\pi$-polarization of the
C(111) reflection from a $0.1$ mm-thick diamond crystal.  The
advantage of using the $\pi$-polarization is due to the fact that
the bandwidth of the reflectance is a few times narrower than for
the $\sigma$-polarization in the entire energy range.  As a result,
the profile of the transmitted intensity in the time-domain appears
more suitable for the temporal windowing operation, especially if
one is interested in nearly Fourier-limited pulses.

While the radiation is sent through the crystal, the electron beam
passes through a magnetic chicane, which accomplishes three tasks by
itself: it creates an offset for the crystal installation, it
removes the electron micro-bunching produced in the first undulator,
and it acts as a delay line for the implementation of the temporal
windowing. In other words, the magnetic chicane shifts the electron
bunch on top of the monochromatic wake created by the bandstop
filter thus selecting a part of the wake. This operation amounts to
a temporal windowing process. By this, the electron bunch is seeded
with a radiation pulse characterized by a bandwidth much narrower
than the natural FEL bandwidth. The temporal shape of the seed
signal is shown in the left plot of Fig. \ref{seed1}, while the
seeded FEL signal before the second self-seeding setup is shown in
the right plot of the same figure.

For the hard X-ray wavelength range, a small dispersive strength
$R_{56}$  in the order of ten microns is sufficient to remove the
micro bunching in the electron bunch. It is important to realize
that the uncorrelated energy spread induced by quantum diffusion
during the passage through the undulator is more than sufficient to
wash-out the microbunching. In our case of interest, the energy
spread induced by quantum diffusion always exceeds $1$
MeV\footnote{In this article we discuss about SASE3 type undulators
placed behind SASE1, which induces a relatively high energy spread
in electron bunch due to quantum diffusion, definitely exceeding  1
MeV at the nominal energy of 17.5 GeV. }. It should be remarked that
this effect is of fundamental nature, and that the energy-spread so
produced follows a Gaussian distribution. Based on this fact, the
energy and density modulations will be damped following an
exponential factor given by $\exp[-<(\delta \gamma)^2> R_{56}^2/(2
\gamma^2 \lambdabar^2)]$, where $<(\delta \gamma)^2>/\gamma^2$ is
the variance of the relative energy spread, $R_{56}$ is the
dispersion strength of the chicane, and $\lambdabar$ is the reduced
wavelength. Considering a wavelength $\lambda = 0.35$ nm, an
electron energy of $17.5$ GeV, an $R_{56} \sim 30 ~\mu$m and,
conservatively, an energy spread induced by quantum diffusion of $1$
MeV, we obtain an enormously small exponential damping factor.

As a result of this discussion, the choice of the strength of the
magnetic chicane only depends on the delay that we want to introduce
between electron bunch and radiation. In our case, this amounts to
$17 ~\mu$m for the short pulse mode of operation. Such dispersion
strength is small enough to be generated by a short $5$ m-long
chicane to be installed in place of a single undulator module. Such
chicane is, however, strong enough to create a sufficiently large
transverse offset of a few millimeters for installing the crystal.

The main problem in having crystal monochromators working between
$3$ keV and $5$ keV is related with the low throughput due to
absorption. In fact, about $88 \%$ of the radiation out of the
reflection range is absorbed, resulting in low seeding power. The
seed power level can be increased, to some extent, by making the
first part of the undulator longer. This increases the
signal-to-noise ratio. However, successful operation of the
self-seeded XFEL requires operation of the first part of the
undulator in the deep linear regime, and not in saturation. In fact,
the amplification process in the FEL leads to an energy modulation
in the electron beam. After the electron beam passes through the
magnetic chicane, such energy modulation transforms into additional
energy spread.

The use of a second monochromator cascade enhances the
signal-to-noise ratio without spoiling the electron beam quality.
This enhancement is the consequence of the fact that the radiation
pulse impinging on the second crystal is nearly Fourier-limited,
meaning that there is an improvement of the bandwidth ratio of the
signals before the first cascade (SASE) and the second cascade
(seeded). This fact can be easily seen by comparing the spectral
widths in Fig. \ref{bio2f3} and Fig. \ref{bio2f4}, resulting in an
increase of about a factor $5$. Note that the power level before the
first and the second cascade are roughly the same, and amount to
about $1$ GW, Fig. \ref{powcomp}. Therefore, the improvement of the
bandwidth ratio directly translates into an in improvement of the
seed power (compare Fig. \ref{seed1} and Fig. \ref{seed2}) of the
same amount. The power level in Fig. \ref{powcomp} is much smaller
than the power at saturation, which reaches about $100$ GW, so that
the electron beam is not significantly perturbed.

This fact is confirmed by simulations. In fact, the seed signal is
finally amplified through the output undulator, which is tapered in
order to optimize the exchange between electron energy and radiation
(see the next section for details). The FEL signal from the entire
setup is shown in the left plot of Fig. \ref{outf}, which also
includes a plot of the energy in the FEL pulse as a function of the
distance inside the output undulator. As one can see, for this
particular run one reaches about 2 TW output power.

If one wants to reach the same seed power with a single cascade, one
needs to increase the power at the exit of the first undulator up to
about 5 GW, thus perturbing the electron beam more. Simulations show
that due to electron beam perturbations, in this case one cannot
reach the same final output level of 2 TW.

\section{FEL studies}

\begin{figure}[tb]
\includegraphics[width=0.5\textwidth]{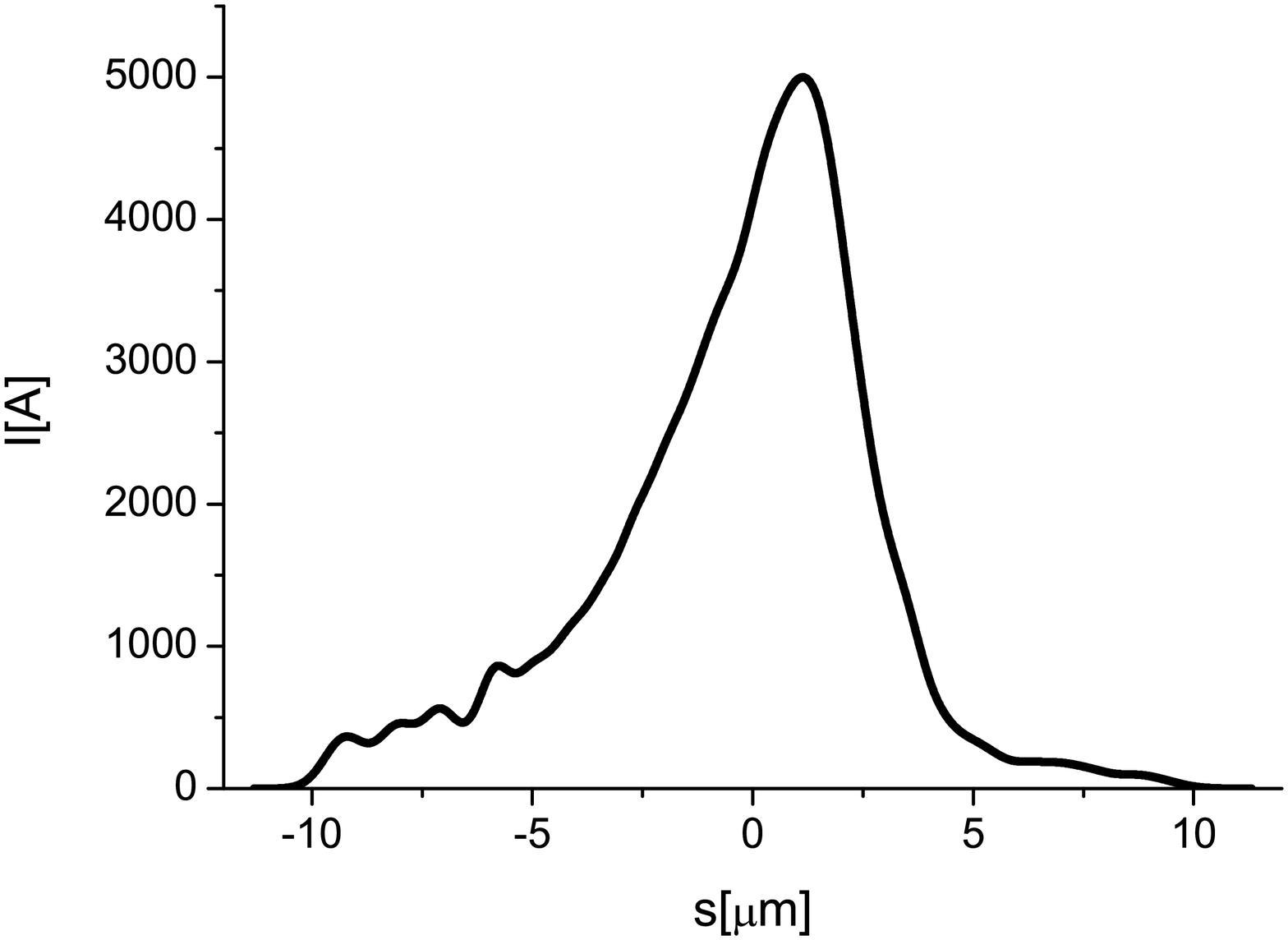}
\includegraphics[width=0.5\textwidth]{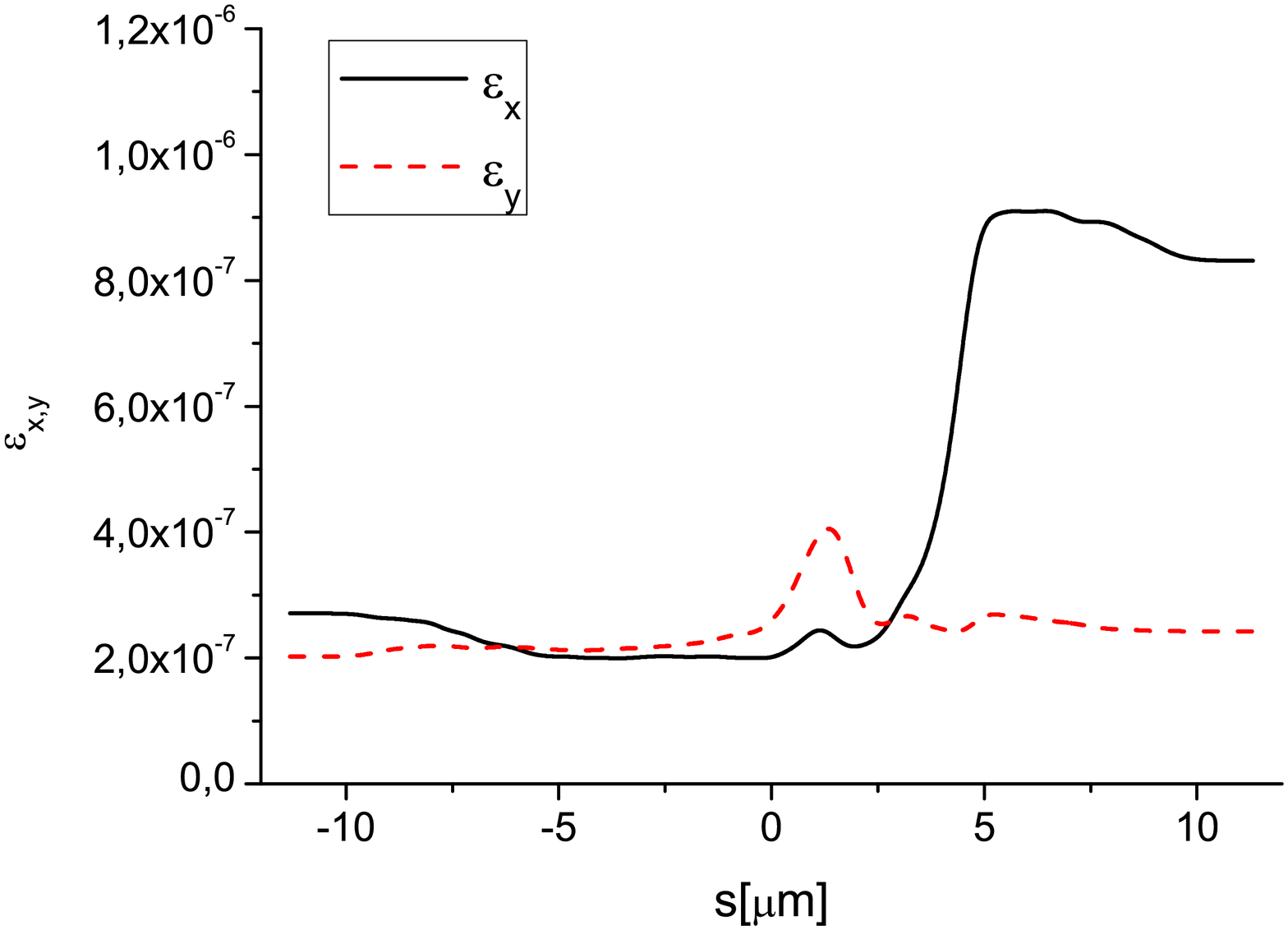}
\includegraphics[width=0.5\textwidth]{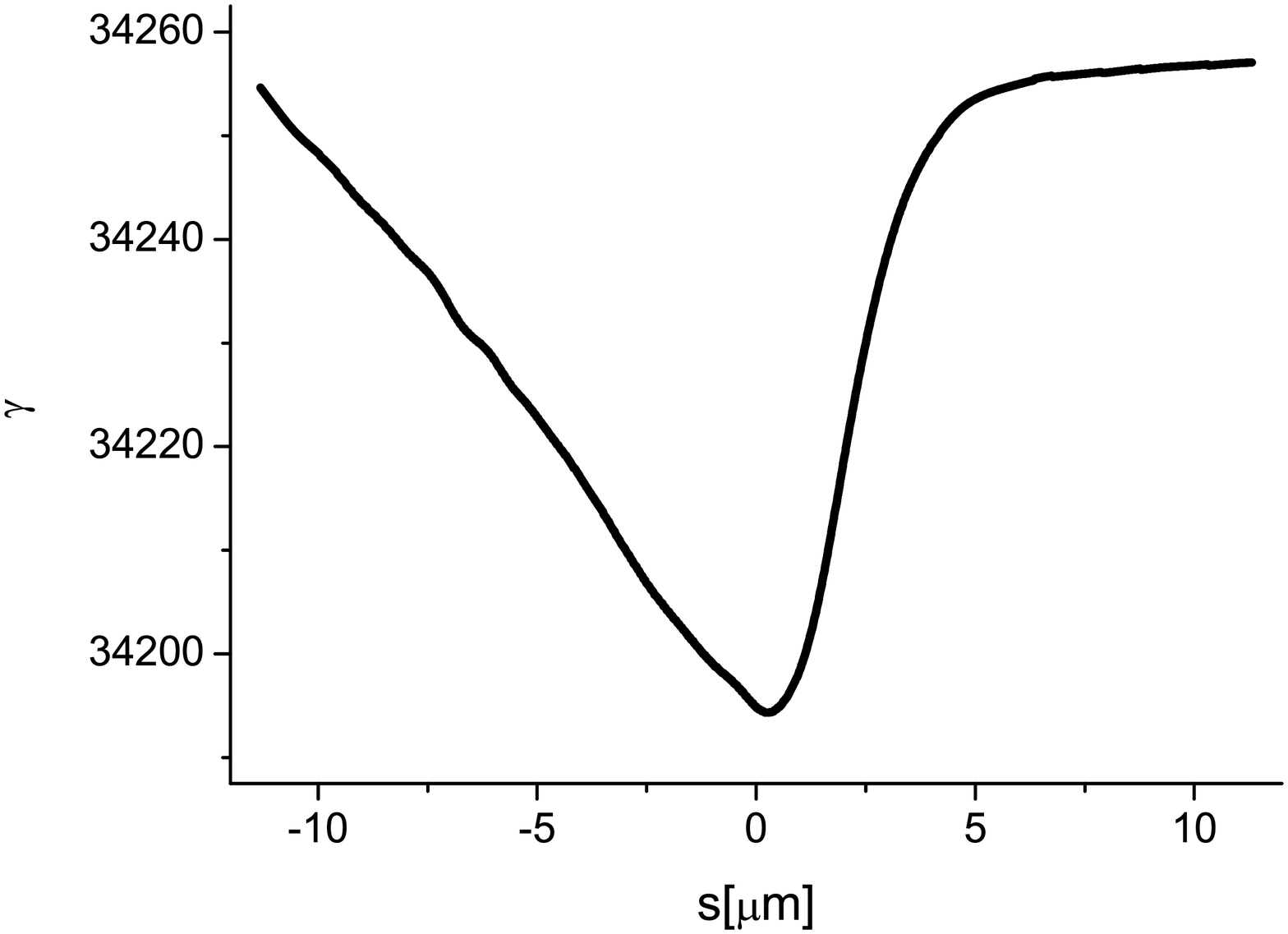}
\includegraphics[width=0.5\textwidth]{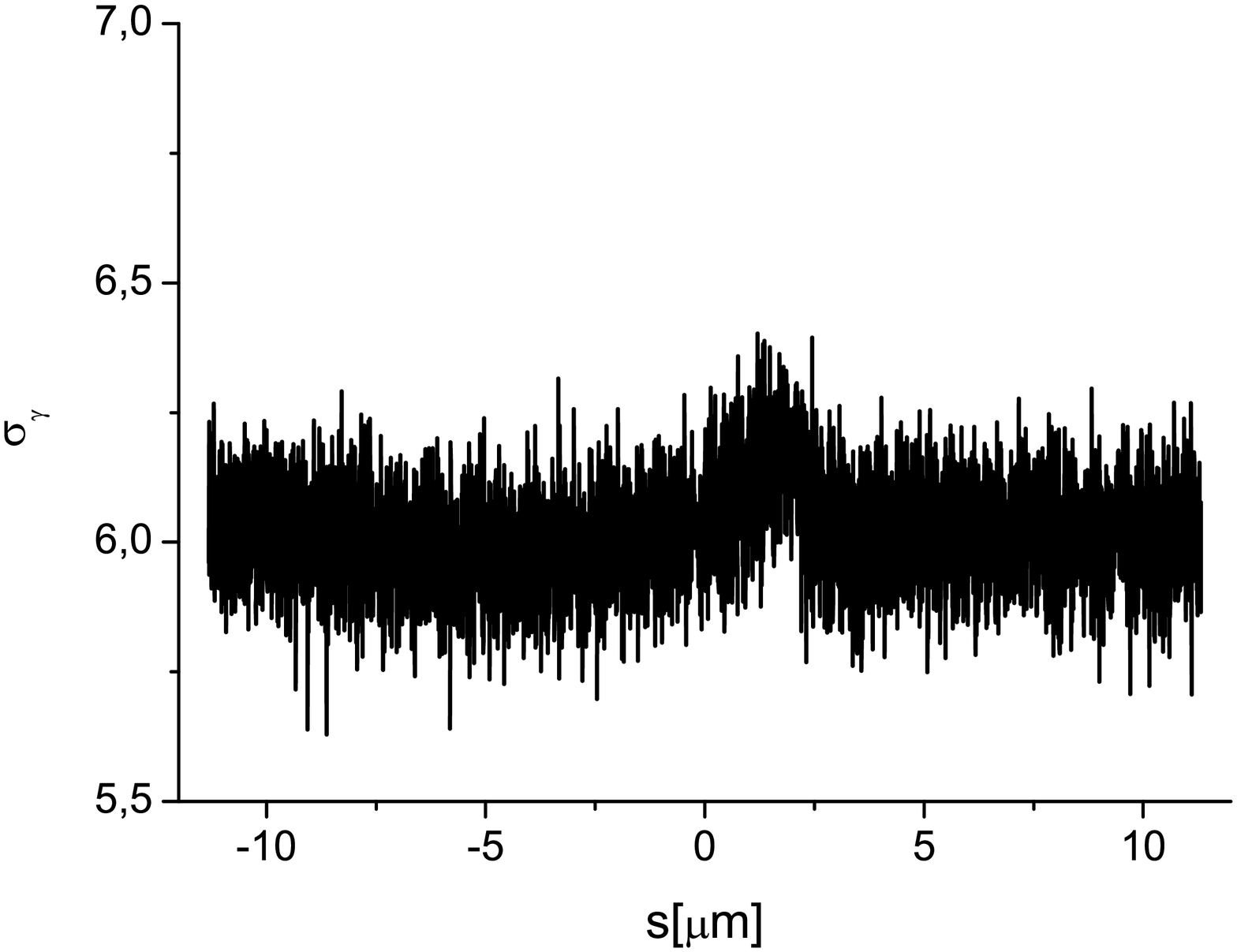}
\begin{center}
\includegraphics[width=0.5\textwidth]{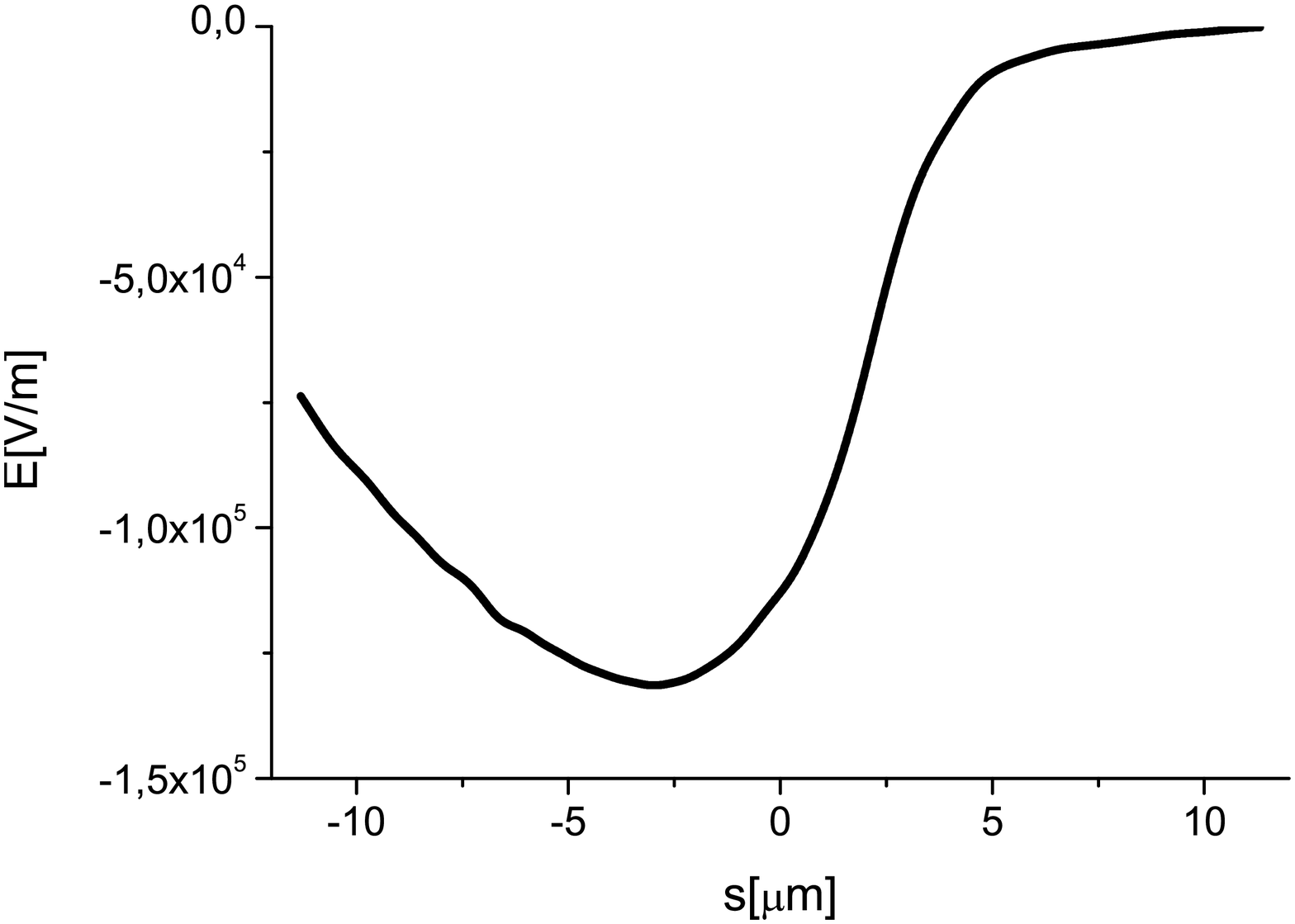}
\end{center}
\caption{Results from electron beam start-to-end simulations at the
entrance of SASE3 \cite{S2ER} for the hard X-ray case. (First Row,
Left) Current profile. (First Row, Right) Normalized emittance as a
function of the position inside the electron beam. (Second Row,
Left) Energy profile along the beam, lower curve. The effects of
resistive wakefields along SASE1 are illustrated by the comparison
with the upper curve, referring to the entrance of SASE1 (Second
Row, Right) Electron beam energy spread profile, upper curve. The
effects of quantum diffusion along SASE1 are illustrated by the
comparison with the lower curve, referring to the entrance of SASE1.
(Bottom row) Resistive wakefields in the SASE3 undulator
\cite{S2ER}.} \label{biof2f3}
\end{figure}

With reference to Fig. \ref{bio2f1} we performed feasibility studies
pertaining the energy range considered in this article. These
studies were performed with the help of the FEL code GENESIS 1.3
\cite{GENE} running on a parallel machine. Simulations are based on
a statistical analysis consisting of $100$ runs.

The main undulator parameters are reported in Table \ref{tt1}. The
electron beam characteristics at the entrance of the setup are
summarized in Fig. \ref{biof2f3}, where we plot the results of
start-to-end simulations \cite{S2ER}.

\begin{table}
\caption{Undulator parameters}

\begin{small}\begin{tabular}{ l c c}
\hline & ~ Units &  ~ \\ \hline
Undulator period      & mm                  & 68     \\
Periods per cell      & -                   & 73   \\
Total number of cells & -                   & 40    \\
Intersection length   & m                   & 1.1   \\
Photon energy         & keV                 & 3-5 \\
\hline
\end{tabular}\end{small}
\label{tt1}
\end{table}

\begin{figure}[tb]
\includegraphics[width=0.5\textwidth]{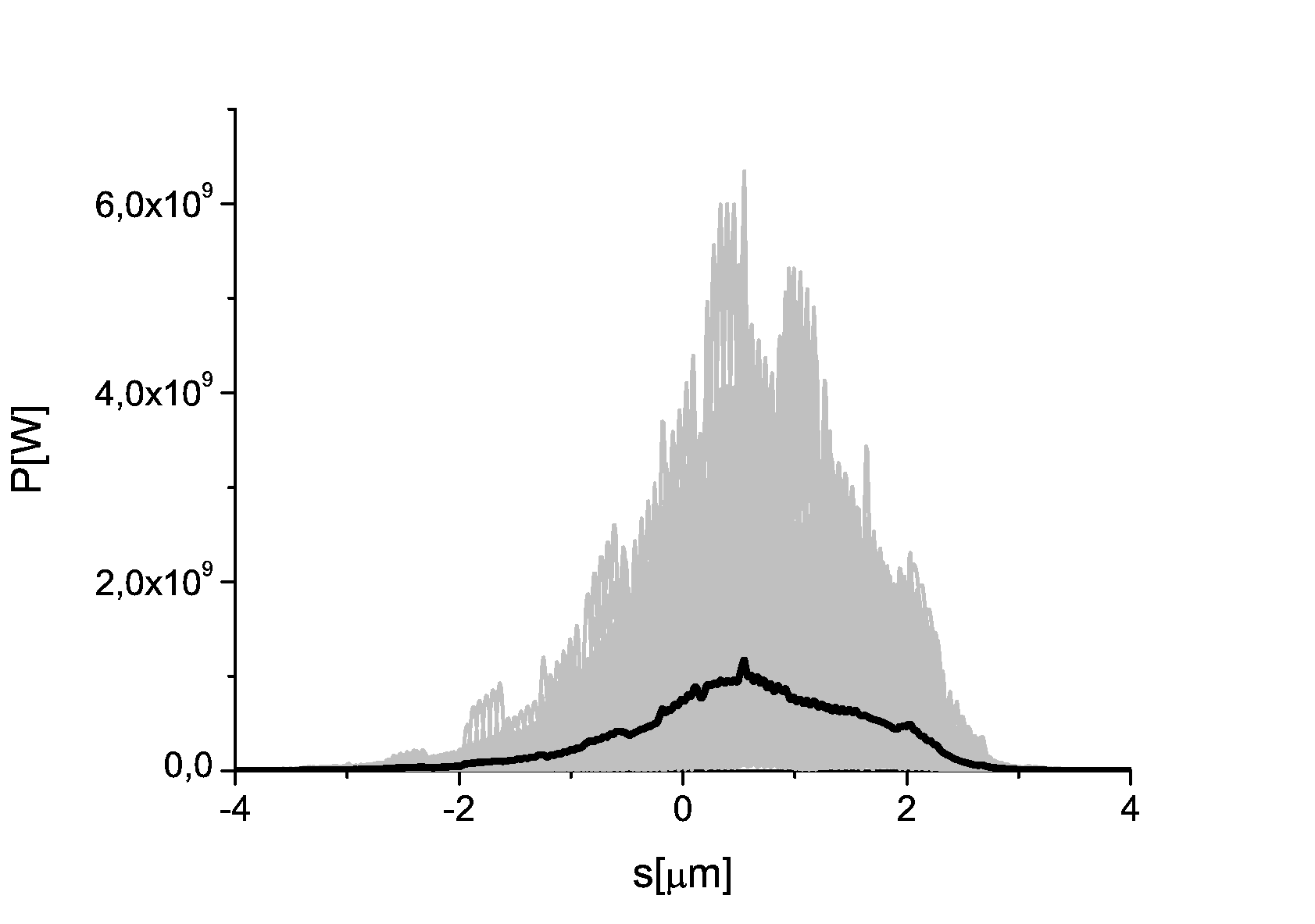}
\includegraphics[width=0.5\textwidth]{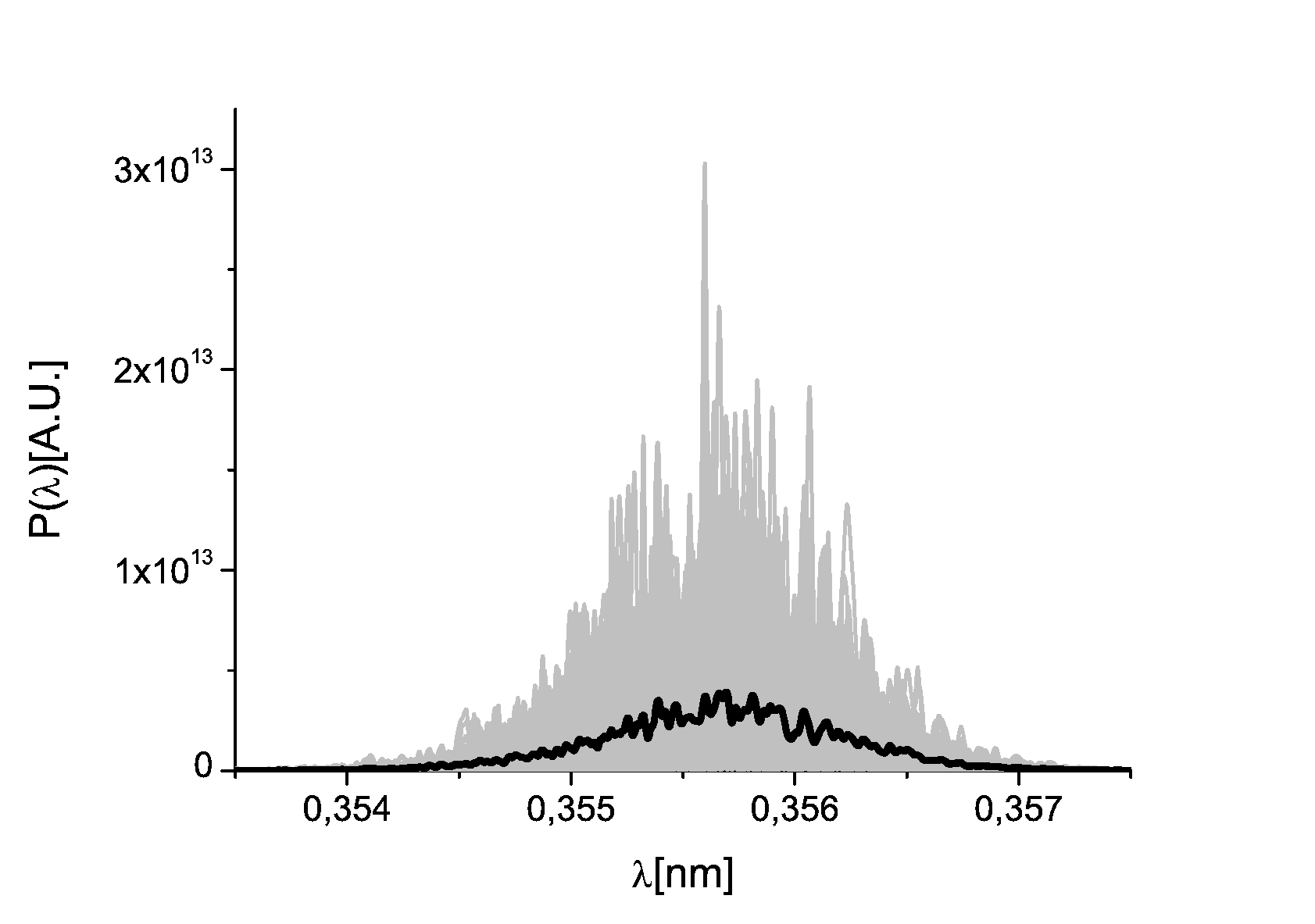}
\caption{Power and spectrum before the first hard X-ray self-seeding
monochromator setup. Grey lines refer to single shot realizations,
the black line refers to the average over a hundred realizations.}
\label{bio2f4b}
\end{figure}
First, the electron beam passes through the first part of the
undulator and lases as a SASE source. The power and the spectrum
after this first undulator, and before the first self-seeding
monochromator setup are shown in Fig. \ref{bio2f4b}. This photon
pulse passes through the first crystal acting as a band-stop filter
as described before.

\begin{figure}[tb]
\includegraphics[width=0.5\textwidth]{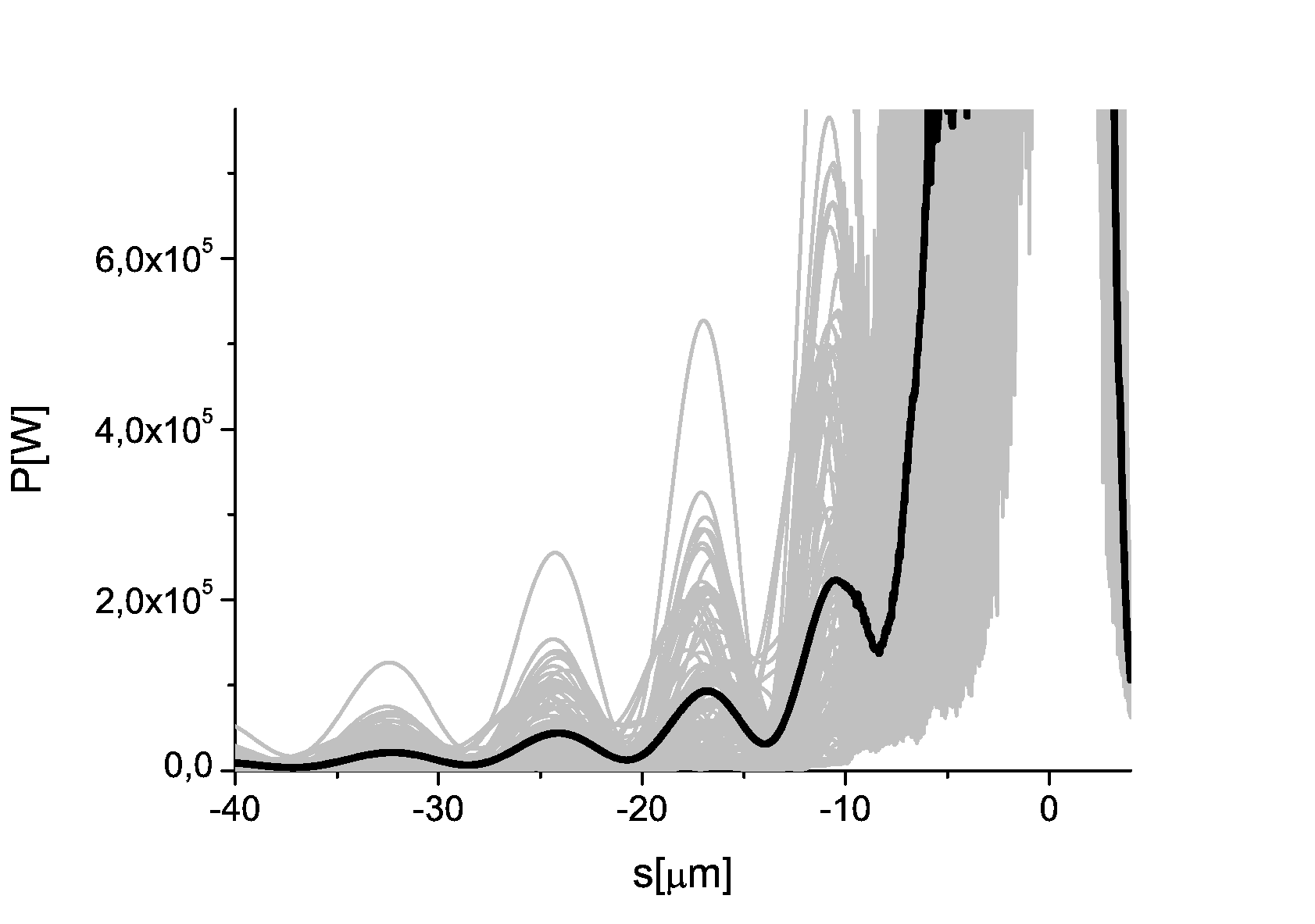}
\includegraphics[width=0.5\textwidth]{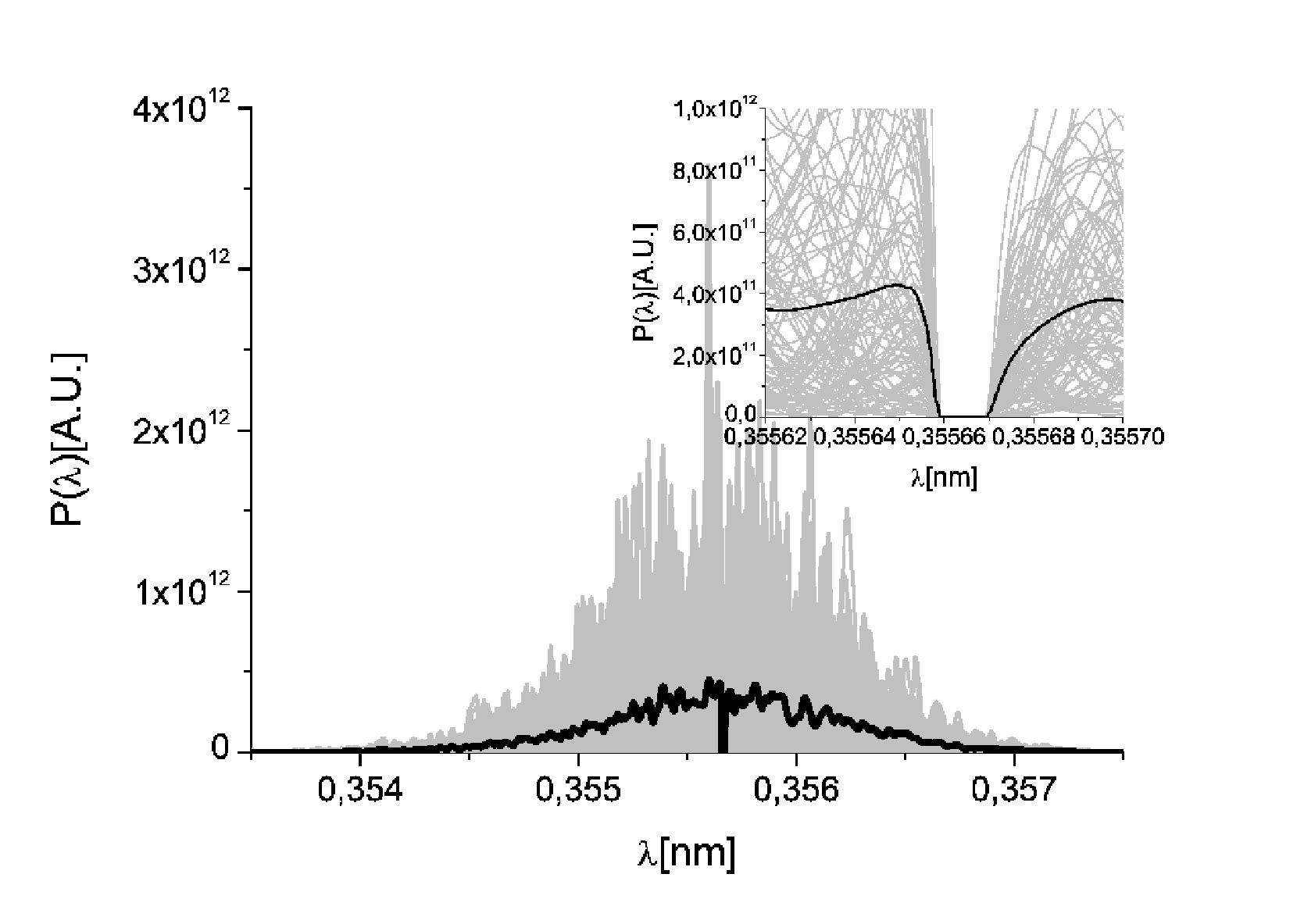}
\caption{Power and spectrum after the first hard X-ray self-seeding
monochromator setup. Grey lines refer to single shot realizations,
the black line refers to the average over a hundred realizations.}
\label{bio2f5}
\end{figure}
The results in frequency and time domain are shown in Fig.
\ref{bio2f5}, where the trailing radiation pulse in the time-domain,
due to the presence of the monochromator, is evident.

\begin{figure}[tb]
\includegraphics[width=0.5\textwidth]{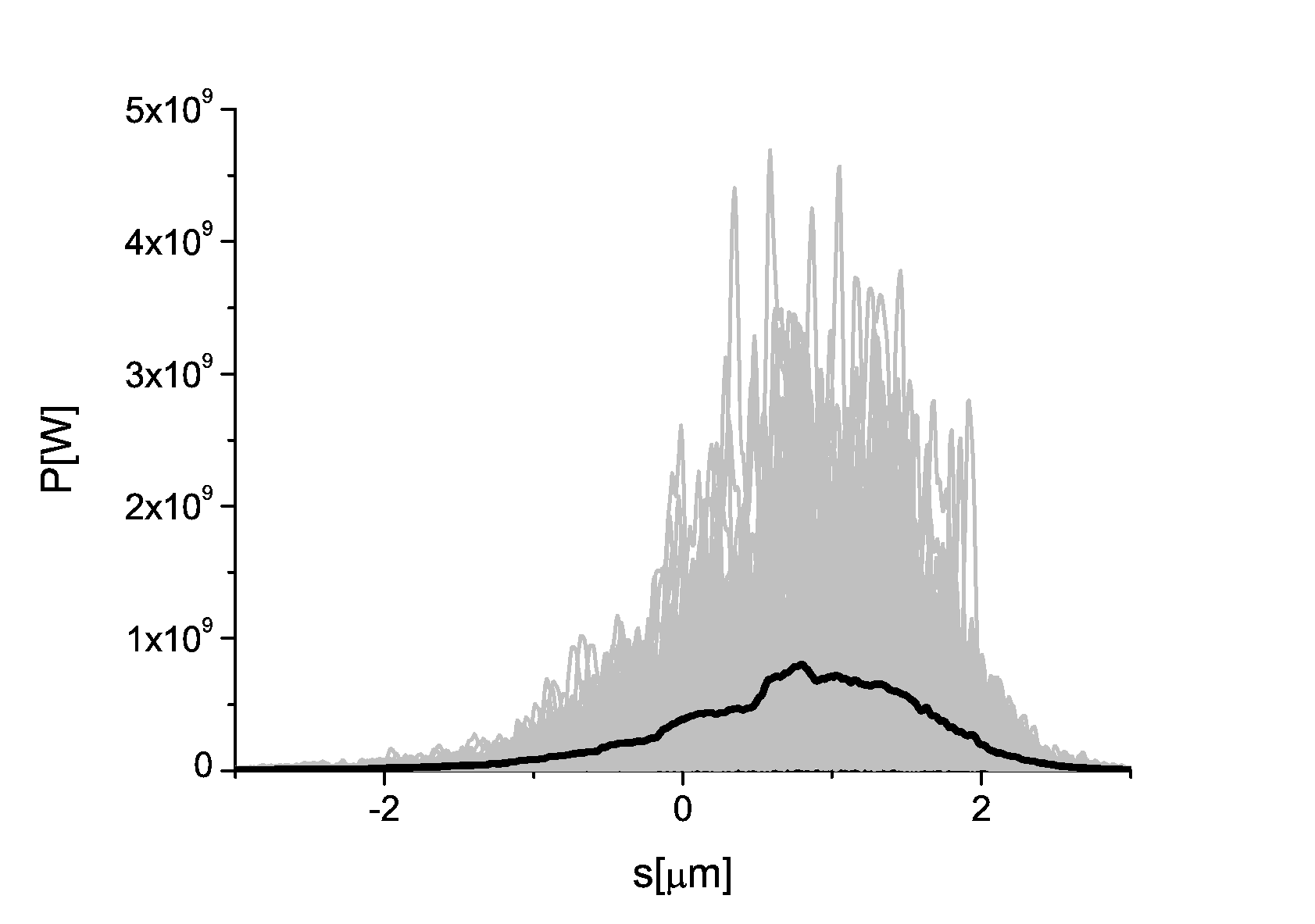}
\includegraphics[width=0.5\textwidth]{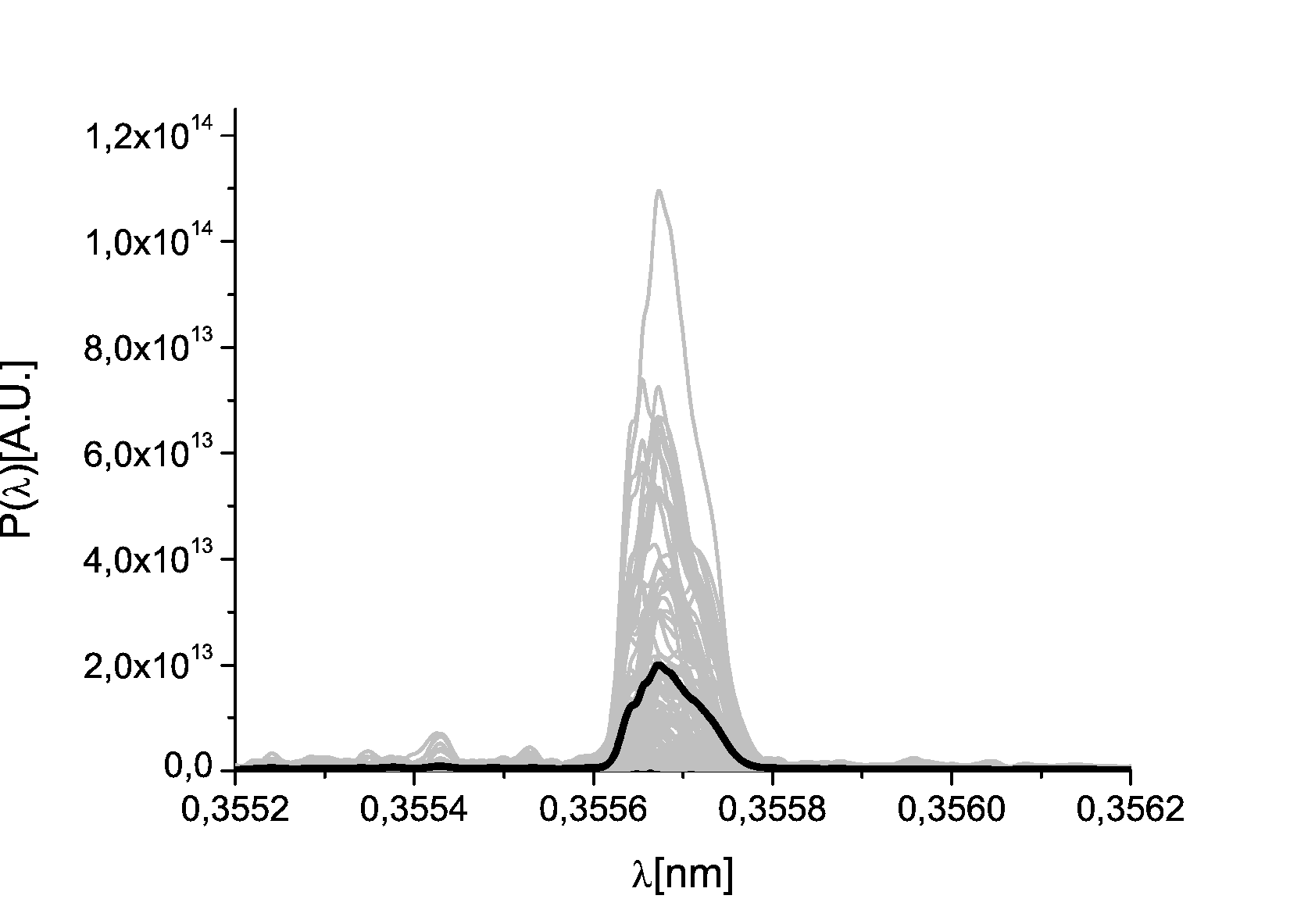}
\caption{Power and spectrum before the second hard X-ray
self-seeding monochromator setup. Grey lines refer to single shot
realizations, the black line refers to the average over a hundred
realizations.} \label{bio2f6}
\end{figure}
\begin{figure}[tb]
\includegraphics[width=0.5\textwidth]{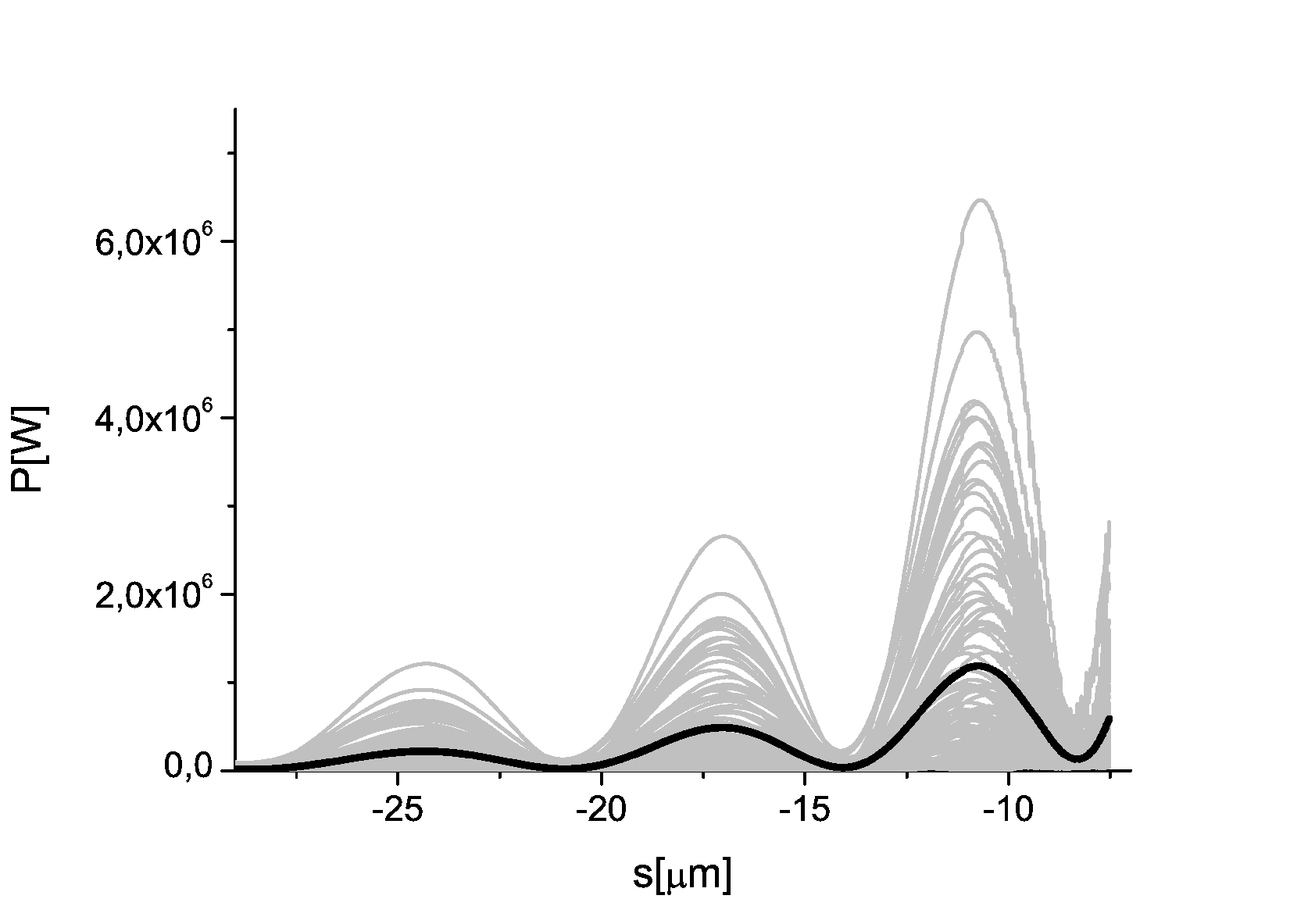}
\includegraphics[width=0.5\textwidth]{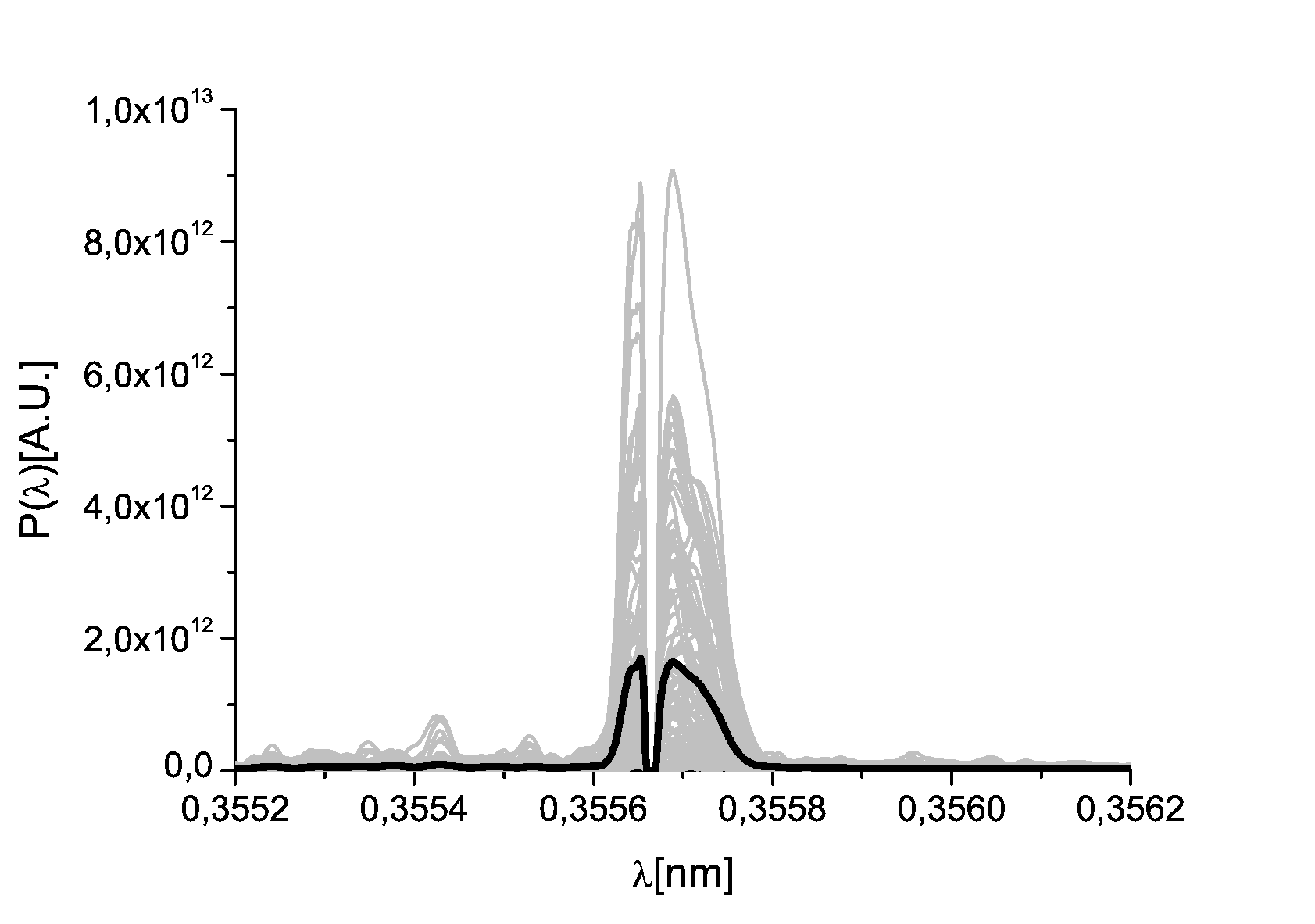}
\caption{Power and spectrum after the second hard X-ray self-seeding
monochromator setup. Grey lines refer to single shot realizations,
the black line refers to the average over a hundred realizations.}
\label{bio2f7}
\end{figure}

The electron beam modulations are washed-out by passing through the
chicane, and is seeded with the trailing radiation pulse, which is
then amplified in the second undulator part. The power and the
spectrum of the radiation after the second part of the undulator is
shown in Fig. \ref{bio2f7}.

At this point the radiation pulse passes through the second
monochromator setup. The fact that it is nearly Fourier limited
allows a betterment of the signal-to-noise ratio of a large factor
$\Delta \omega_{\mathrm{SASE}} \cdot \sigma_T$. This helps in
countering the high absorption in the crystal. Fig. \ref{bio2f7}
shows the power and the spectrum after the second monochromator
station.

\begin{figure}[tb]
\begin{center}
\includegraphics[width=0.5\textwidth]{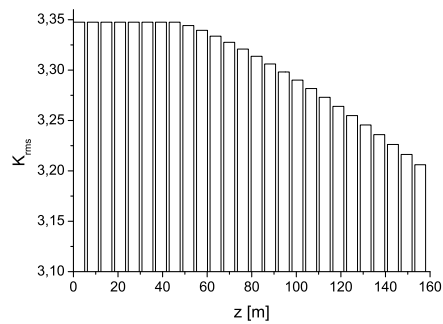}
\end{center}
\caption{Tapering law.} \label{taplaw}
\end{figure}

\begin{figure}[tb]
\includegraphics[width=0.5\textwidth]{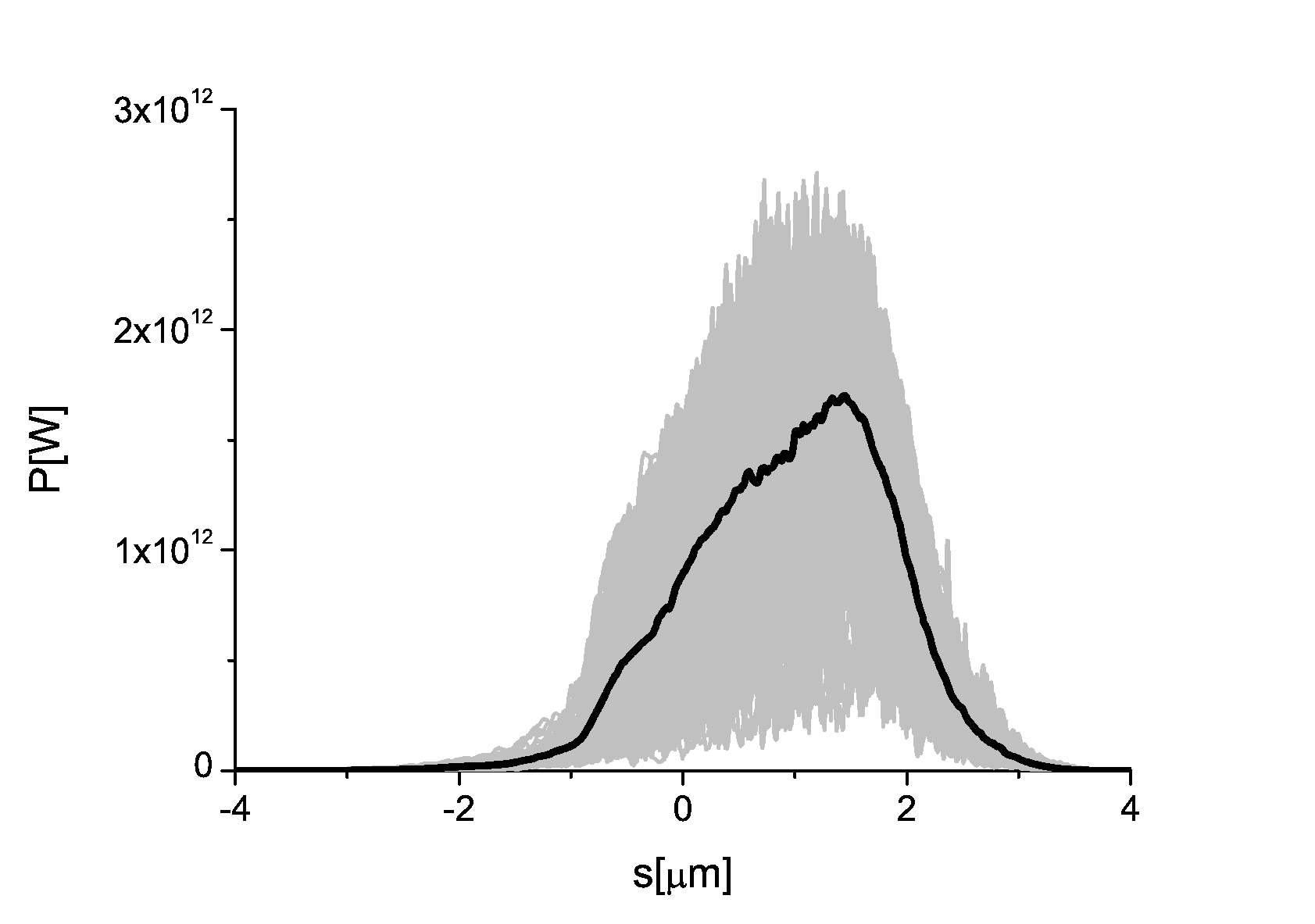}
\includegraphics[width=0.5\textwidth]{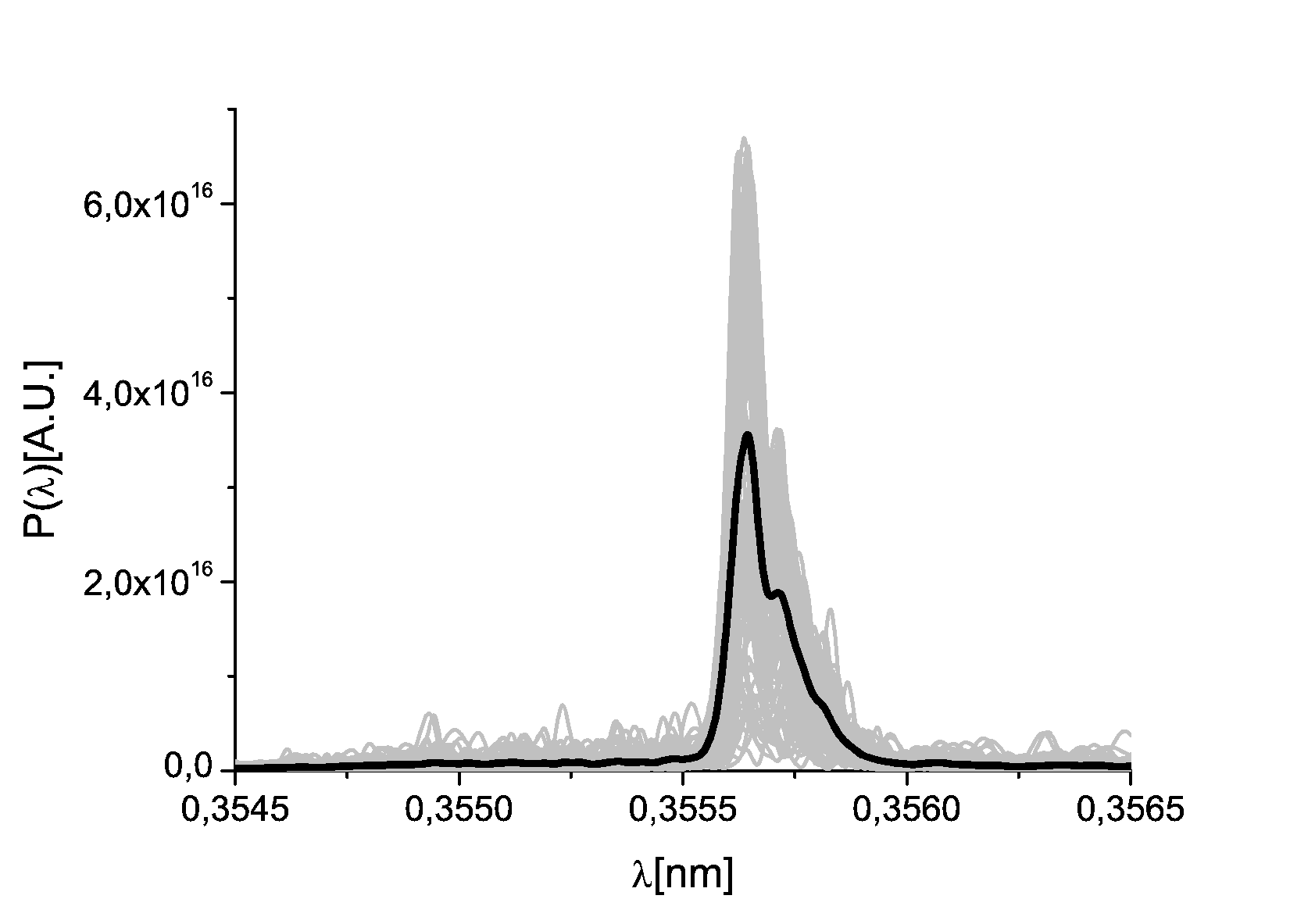}
\caption{Final output in the case of tapered output undulator for
$\lambda = 0.35$ nm. Power and spectrum are shown. Grey lines refer
to single shot realizations, the black line refers to the average
over a hundred realizations.} \label{biofh8}
\end{figure}

\begin{figure}[tb]
\includegraphics[width=0.5\textwidth]{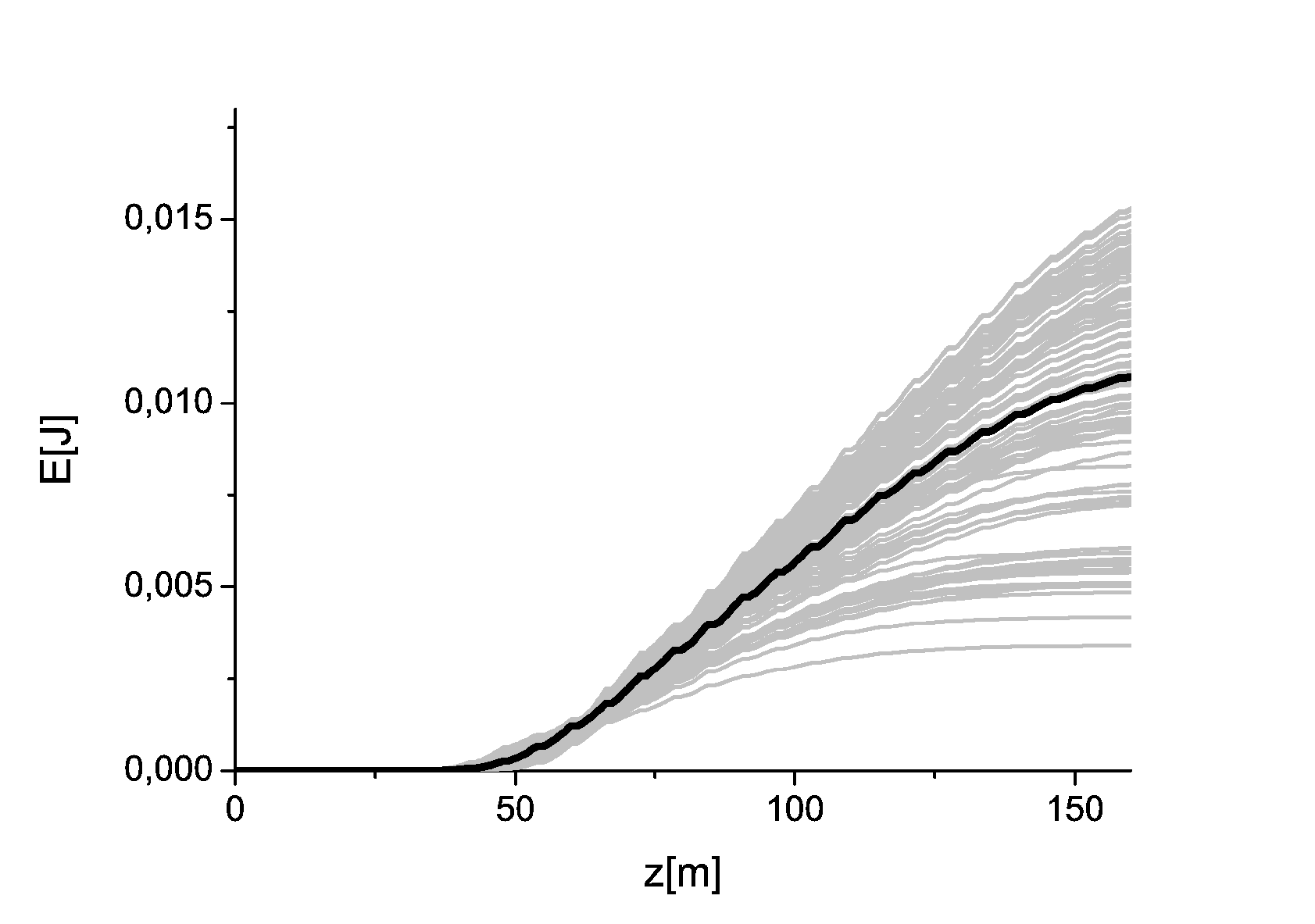}
\includegraphics[width=0.5\textwidth]{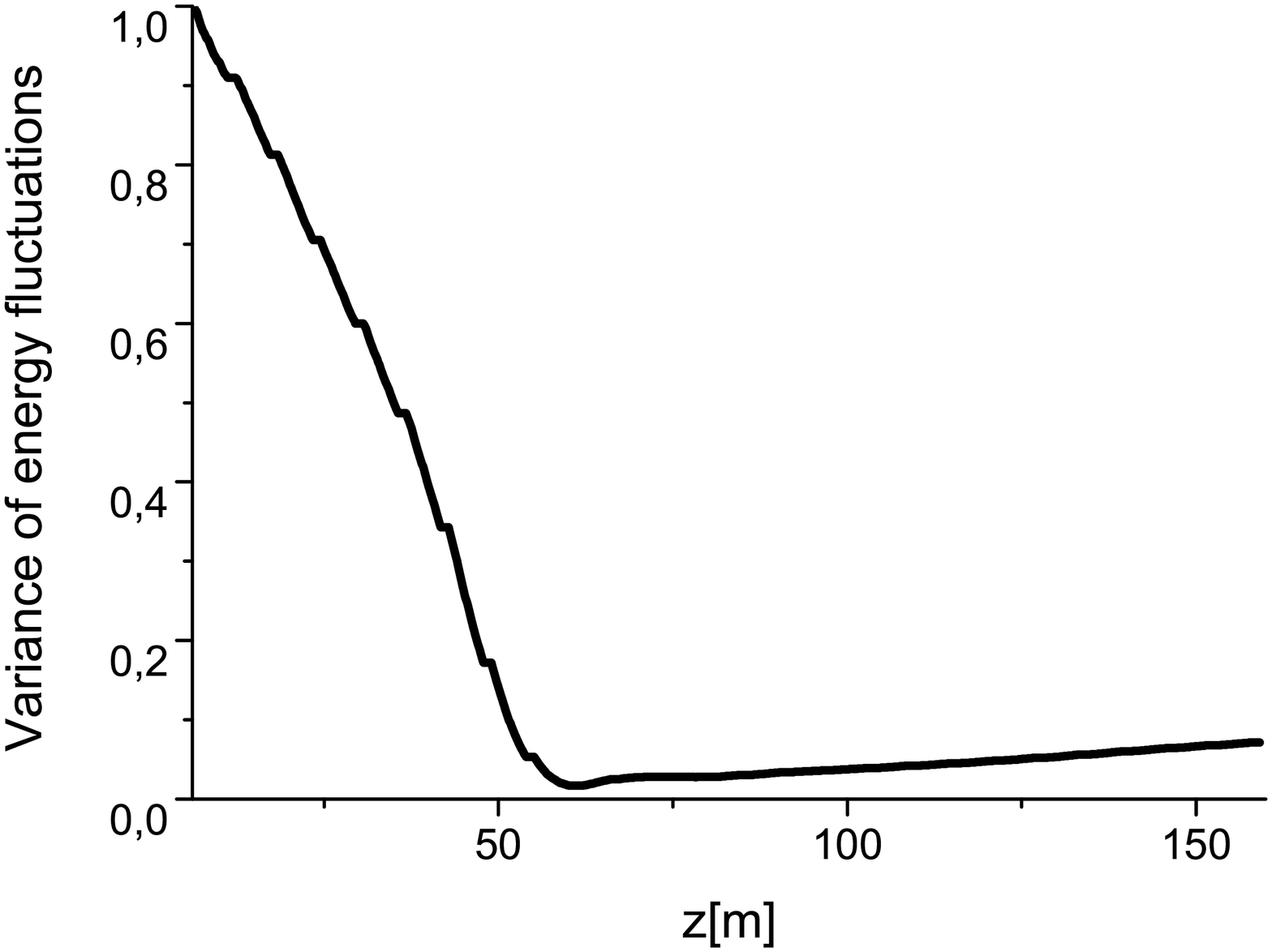}
\caption{Energy and energy variance of output pulses in the case of
tapered output undulator for $\lambda = 0.35$ nm. In the left plot,
grey lines refer to single shot realizations, the black line refers
to the average over a hundred realizations.} \label{biofh9}
\end{figure}

\begin{figure}[tb]
\includegraphics[width=0.5\textwidth]{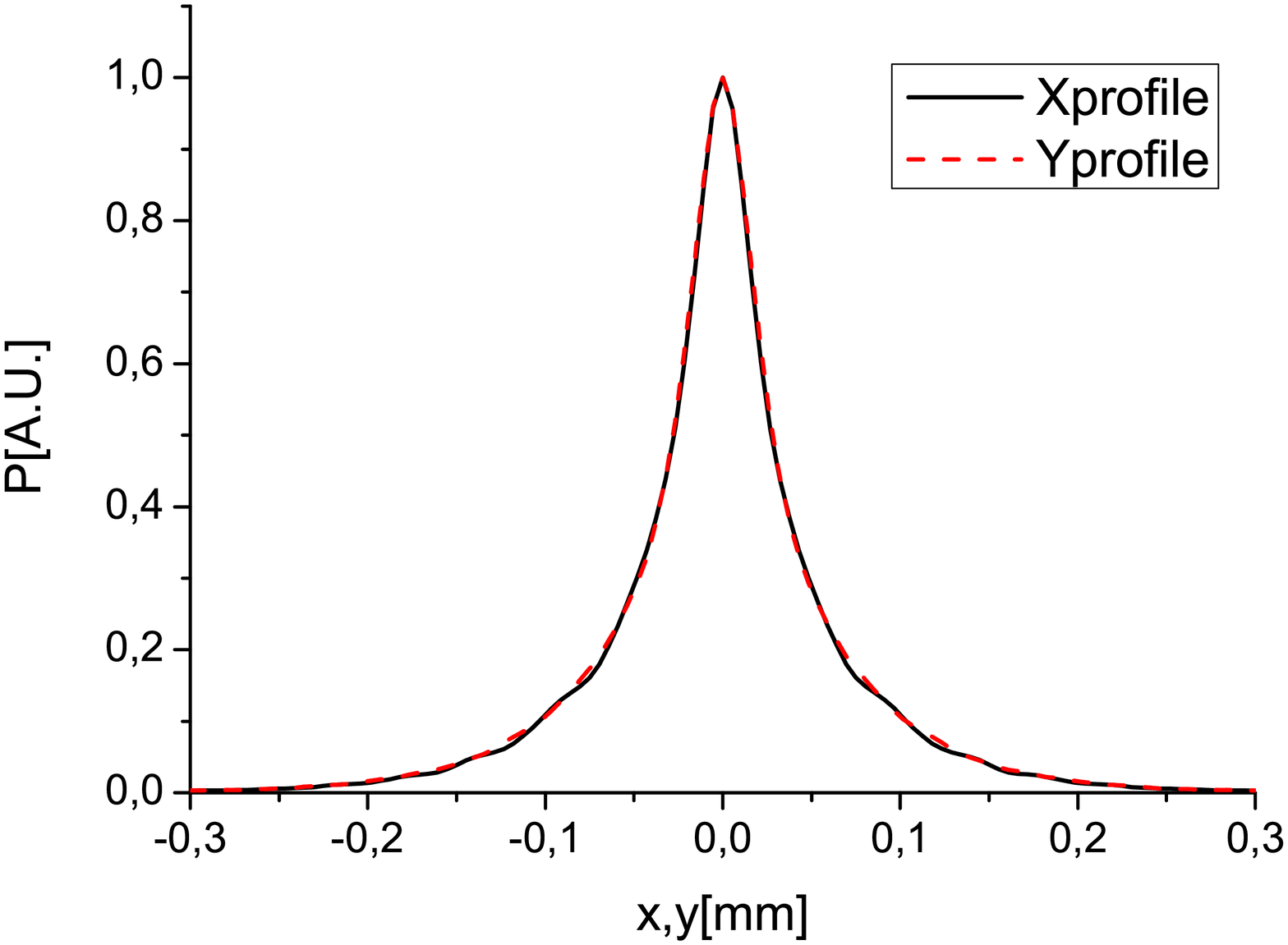}
\includegraphics[width=0.5\textwidth]{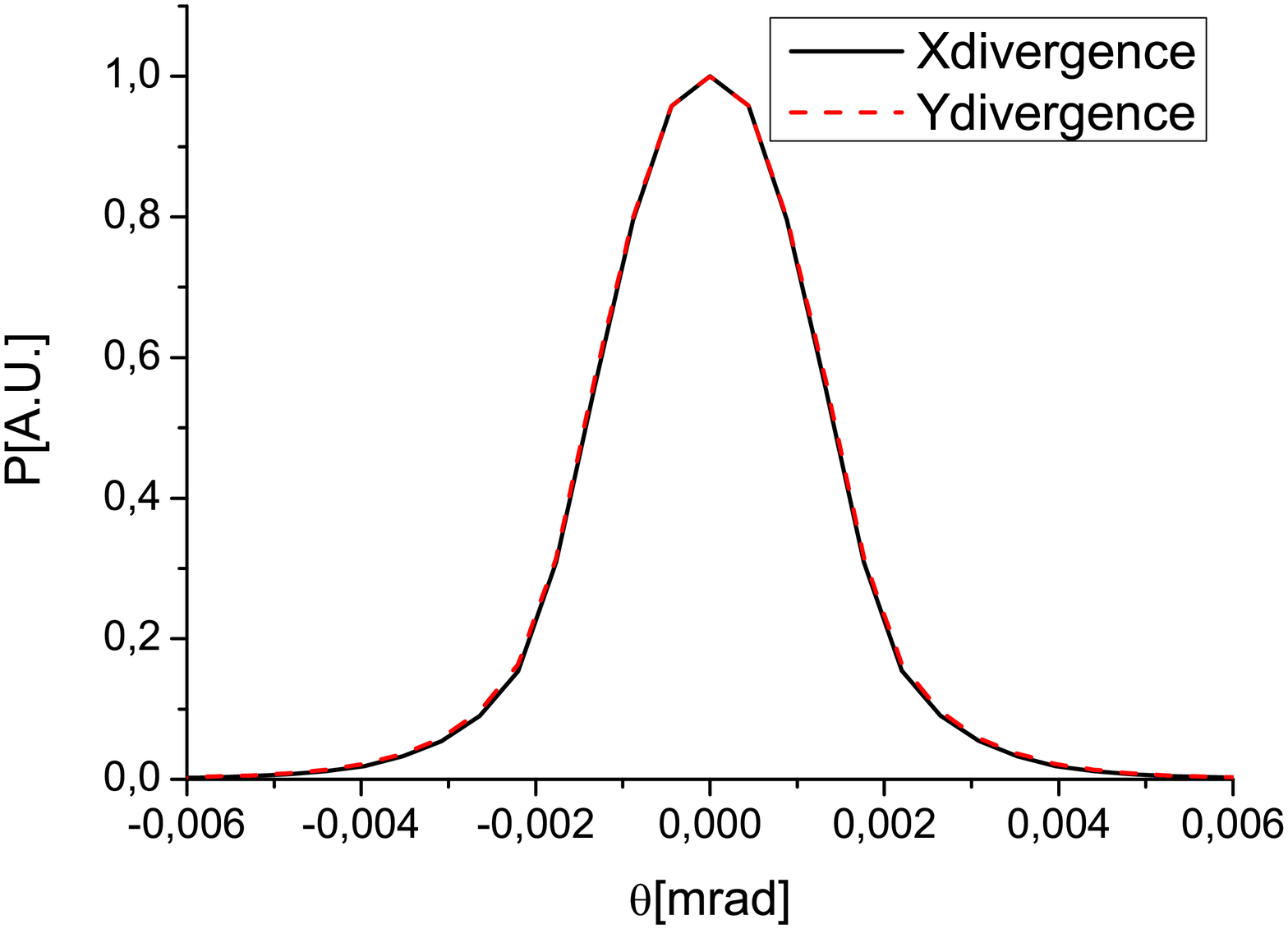}
\caption{Final output. X-ray radiation pulse energy distribution per
unit surface and angular distribution of the X-ray pulse energy at
the exit of output undulator for the case $\lambda = 0.35$ nm.}
\label{biofh10}
\end{figure}

The electron beam finally goes through the output undulator, which
is tapered according to the law shown in Fig. \ref{taplaw}. Fig.
\ref{biofh8}, Fig. \ref{biofh9} and Fig. \ref{biofh10} show the
output from the entire setup. Fig. \ref{biofh8} demonstrates that
nearly Fourier-limited pulses with a power level of about 2 TW can
be reached by tapering the output undulator. Fig. \ref{biofh9} shows
the energy and the variance of the radiation pulses as a function of
the position inside the undulator. Finally, size and divergence of
the X-ray pulses are plotted in Fig. \ref{biofh10}.

\section{Conclusions}

A self-seeding scheme based on single crystal monochromator has been
demonstrated experimentally and successfully compared with
simulations \cite{EMHX}. Self-seeding is an excellent method for
generating both monochromatic X-rays and high power pulses. Up to
now, all studies have focused on operation in the hard X-ray
energies (between 7 keV and 13 keV). However, future setups may need
to be operated at lower energies. The interest in biomolecular
imaging now includes also sources between 3 keV and 5 keV, and
self-seeding schemes began to cope with this energy range
\cite{OBIO}. In this paper we demonstrated the flexibility of our
self-seeding scheme with single crystal monochromator to cover a
wide photon energy range. This kind of operation is easily achieved
with diamond crystals in symmetric Bragg reflection geometry. In
particular, based on the use C(400), C(220), and C(111) reflections
it will be possible to cover photon energy range from 13 keV down to
3.5 keV. Therefore, this extremely compact self-seeding scheme is
ideally suited for bio-imaging applications. In this paper we
proposed a study of the performance of the self-seeding scheme with
single crystal monochromator for the European X-ray FEL at X-ray
energies lower than 5 keV. By combining the two techniques of
cascade self-seeding and undulator tapering we found that 2 TW
X-ray, nearly transform-limited pulses down to photon energy range
3.5 keV can be generated from baseline-scale $40$ cells undulators.

\section{Acknowledgements}

We are grateful to Massimo Altarelli, Reinhard Brinkmann, Henry
Chapman, Janos Haidu, Viktor Lamzin, Serguei Molodtsov and Edgar
Weckert for their support and their interest during the compilation
of this work.

\end{document}